\documentclass[sigconf,nonacm,9pt]{acmart}
\usepackage[utf8]{inputenc}
\usepackage[T1]{fontenc}

\usepackage{algorithmic}
\usepackage[linesnumbered,ruled]{algorithm2e}
\usepackage{subcaption}
\usepackage{pifont}
\usepackage{overpic}
\usepackage{soul}
\newcommand{\cname}{\emph{Selene}\xspace}
\definecolor{highLightChange}{RGB}{0,0,0}
\definecolor{revised}{RGB}{0,0,0}

\AtBeginDocument{%
  \providecommand\BibTeX{{%
    \normalfont B\kern-0.5em{\scshape i\kern-0.25em b}\kern-0.8em\TeX}}}

\definecolor{yxw}{RGB}{0, 0, 0}
\newcommand{\yx}[1]{\textcolor{yxw}{#1}}

\setcopyright{none}

\begin{document}

\title{\textcolor{revised}{Towards} High-Speed Passive Visible Light Communication with Event Cameras and Digital Micro-Mirrors }
\acmSubmissionID{777}
\settopmatter{printfolios=false}
\settopmatter{printacmref=false,printfolios=true}
\renewcommand{\authors}{Yanxiang Wang, Yiran Shen, Kenuo Xu, Guangrong Zhao,  Mahbub Hassan, Cheren Xu and Wen Hu}

\author{Yanxiang Wang}

\orcid{0000-0002-1466-4006}
\affiliation{%
  \institution{University of New South Wales}
  \institution{CSIRO}
  \country{Australia}
}
\email{yanxiang.wang@unsw.edu.au}

\author{Yiran Shen}
\orcid{0000-0003-1385-1480}
\affiliation{%
  \institution{Shandong University}
  \country{China}
}
\email{yiran.shen@sdu.edu.cn}
\authornote{Yiran Shen is the corresponding author}
\author{Keruo Xu}
\orcid{0000-0003-3774-644X}
\affiliation{%
  \institution{Peking University}
  \country{China}
}
\email{kenuo.xu@pku.edu.cn}
\author{Mahbub Hassan}
\orcid{0000-0002-3417-8590}

\affiliation{%
  \institution{University of New South Wales}
  \country{Australia}
}
\email{mahbub.hassan@unsw.edu.au}
\author{Guangrong Zhao}
\orcid{0000-0002-4703-9397}
\affiliation{%
  \institution{Shandong University}
  \country{China}
}
\email{guangrong.zhao@sdu.edu.cn}

\author{Chenren Xu}
\orcid{0000-0001-9171-2596}

\affiliation{%
  \institution{Peking University}
  \country{China}
}
\email{chenren@pku.edu.cn}
\author{Wen Hu}
\orcid{0000-0002-4076-1811}
\affiliation{%
  \institution{University of New South Wales}
  \country{Australia}
}
\email{wen.hu@unsw.edu.au}
\renewcommand{\shortauthors}{Wang, et al.}

\begin{abstract}

Passive visible light communication (VLC) modulates light propagation or reflection to transmit data without directly modulating the light source. Thus, passive VLC provides an alternative to conventional VLC, enabling communication where the light source cannot be directly controlled. There have been ongoing efforts to explore new methods and devices for modulating light propagation or reflection. The state-of-the-art has broken the 100 kbps data rate barrier for passive VLC by using a digital micro-mirror device (DMD) as the light modulating platform, or transmitter, and a photo-diode as the receiver. We significantly extend this work by proposing a massive spatial data channel framework for DMDs, where individual channels can be decoded in parallel using an event camera at the receiver. For the event camera, we introduce event processing algorithms to detect numerous channels and decode bits from individual channels with high reliability. Our prototype, built with off-the-shelf event cameras and DMDs, can decode up to 
$\sim$2,000 parallel channels, achieving a data transmission rate of 1.6 Mbps, markedly surpassing current benchmarks by 16x.

\end{abstract}

\ccsdesc[500]{Hardware~Wireless devices}
\ccsdesc[500]{Computer systems organization~Embedded systems}

\keywords{Visible Light Communication, Event Camera, Passive VLC}

\maketitle

\vspace{-1mm}
\section{introduction}

Visible Light Communication (VLC), which utilizes the license-free visible spectrum to transmit data, has emerged as a promising solution to address the spectrum crisis facing next-generation wireless communication. Conventional VLC systems typically involve active modulation of light sources, such as light-emitting diodes (LEDs), at very high speeds, enabling data rates of hundreds of Mbps\cite{zhou2023design}. While these active VLC systems achieve impressive data rates, they are not suitable for scenarios where modifying the existing lighting infrastructure is challenging, such as in historical or protected buildings, art galleries, museums, archaeological sites, and similar locations.

Recently, there has been a growing interest in a novel VLC paradigm known as passive VLC~\cite{xu2017passivevlc, xu2022exploiting}. Unlike conventional VLC, passive VLC does not rely on the active modulation of light sources for data transmission. Instead, it transmits data by modulating the light propagation or reflection to the receiver without manipulating the light source. This approach opens up the possibility of using any available lighting infrastructure for VLC without the need for specialized emission sources.

Despite its advantages, achieving high data rates with passive VLC poses significant challenges. Unlike conventional VLC systems, where LEDs can be turned on and off at very high speeds to encode data, passive VLC relies on modulating an intermediate device, such as a liquid crystal display (LCD), to modulate light propagation. The physical and mechanical properties of these devices often introduce inherent latencies, which further affect the maximum achievable data rate. For example, until recently, the maximum data rate achievable in passive VLC using LCD modulation was only 1 kbps~\cite{ghiasi2021principled}. In a groundbreaking work called PhotoLink~\cite{xu2022exploiting}, researchers set a new benchmark in passive VLC by achieving a data transmission rate of 100 kbps at a short range of less than 2 meters. This was made possible by leveraging a unique type of programmable reflective surface known as a Digital Micro-mirror Device (DMD), where the micro-mirrors can be turned on and off at a high frequency rate.

\begin{figure}[tp!]
    \centering
    \includegraphics[width=\columnwidth]{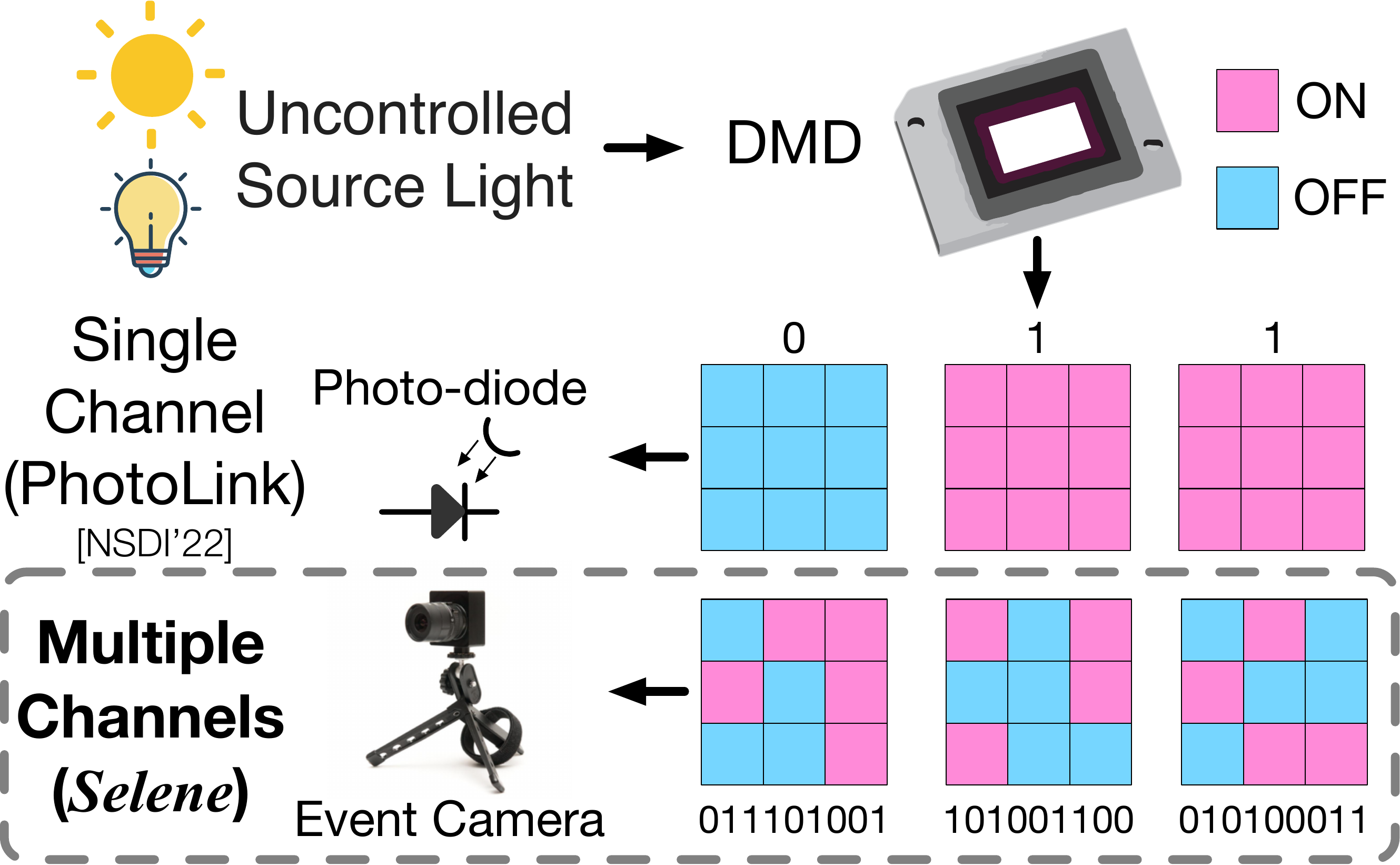}
    \caption{Unlike PhotoLink~\cite{xu2022exploiting}, which configures all micro-mirrors in the DMD to the same state to create a single channel, \cname establishes multiple concurrent VLC channels by controlling clusters of mirrors independently. Moreover, the event camera is selected for its superior temporal resolution compared to conventional RGB cameras and its higher spatial resolution relative to photo-diodes.}
    \label{fig:mchannel}
\end{figure}

While PhotoLink marks a significant milestone in the pursuit of higher data rates in passive VLC, it still falls short of the data rates achievable with conventional VLC. To further boost the data throughput of DMD-based passive VLC, we propose an innovative framework that divides the DMD's entire micro-mirror array into many clusters, as illustrated in Fig.~\ref{fig:mchannel}. Each cluster acts as an independent light-reflecting channel, thereby linearly scaling the total capacity with the number of channels. We propose using an event camera as the receiver to detect and process each individual channel separately. This approach allows for the scaling of data rates using existing DMDs without necessitating complex device customizations to enhance their mirror switching rates, as was the case with PhotoLink. 

\begin{figure}[tp!]
    \centering
    \includegraphics[width=\columnwidth]{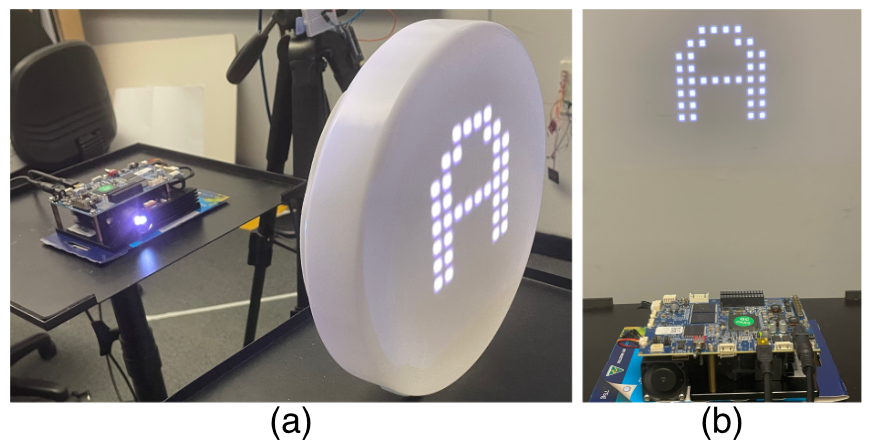}
    \caption{The VLC channels can be distributed in a visual shape (character A) and projected to on light cover cases (a) and ground/walls (b).}
    \label{fig:application}
\end{figure}

Due to the small size (around 7.6 $\mu$m) of the mirrors in DMDs, the light beams from individual mirrors will overlap if the receiver is placed directly in the line of sight of the DMD. This issue can be mitigated by using an intermediate surface, such as walls, light covers, or other objects, between the DMD transmitter and receiver. The channel distribution could exhibit typical visual information, such as logos, while simultaneously transmitting data through VLC channels imperceptible to the human eye, as shown in Fig.~\ref{fig:application}. Such passive VLC can be utilized in tourist locations, museums, and restaurants to deliver maps, historical information, and menu data to users. Additionally, other interesting applications of this technology may emerge in the future that we cannot immediately foresee.

A significant challenge in implementing the proposed multi-channel DMD-based passive VLC lies in effectively differentiating the channels at the receiver's end. To address this, we introduce VLC receivers equipped with event cameras~\cite{gallego2020event}, which excel in detecting rapid changes and pinpointing specific regional activities within the field of view more efficiently than traditional cameras. We have designed event processing algorithms for event cameras that can effectively detect and decode a large number of channels concurrently. Our prototype, constructed with off-the-shelf event cameras and DMDs, achieves a remarkable data transmission rate of 1.6 Mbps, significantly outperforming the established PhotoLink benchmarks by a factor of 16. Surpassing the 1 Mbps threshold is anticipated to facilitate new applications, such as video transmissions, through passive VLC.

The contributions of this paper can be summarized as follows:
\begin{itemize}
    \item To our knowledge, this work represents the first integration of an event camera and a DMD to realize a multi-channel VLC system, called \cname\footnote{\emph{Selene} emanates from the Greek noun \emph{selas}, denoting ``light, brightness, gleam".}, which achieves a data rate of 1.6 Mbps, marking a 16-fold increase over the current state-of-the-art passive VLC systems, such as PhotoLink~\cite{xu2022exploiting}.
    
    \item We propose a rapid channel identification method for event cameras, capable of simultaneously identifying thousands of individual VLC channels in less than two seconds during the channel mapping phase.

    \item We introduce a relative-time decoding method for transmitted bits, designed to mitigate the impact of inaccurate timestamps caused by the asynchronous pixel readout scheme in event cameras. This approach outperforms the conventional absolute-time method, achieving significantly lower bit error rates.

    \item We have designed a dual-rate data encoder for the DMD, specifically to address event cameras' tendency to prioritize central screen events over those on the periphery. This encoder achieves a 5\% improvement in data rates compared to conventional single-rate encoding.

    \item We have implemented the proposed multi-channel VLC system using off-the-shelf event cameras and commodity DMD achieving a \textbf{record data rate of 1.6 Mbps} over a distance of 1m with a Bit Error Rate (BER) under 1\%. 
    \textcolor{revised}{This represents a significant advancement towards high-speed passive VLC.} Further system performance evaluations across diverse conditions demonstrate that our system maintains low BERs up to a range of six meters and exhibits robustness against environment light interference.
      
\end{itemize}

\section{Features of Event Cameras}
\label{EVCameraFeatures}

\label{sec:receiver}
In this section, we highlight key characteristics of event cameras that are crucial for designing a dependable VLC communication system utilizing a DMD transmitter and an event camera receiver. Initially, we discuss the reasons for preferring event cameras over traditional camera technologies.
\subsection{Rationale for Choosing Event Camera}
\label{subsubsec:designChoice}
\begin{figure}[htp!]
    \begin{subfigure}[b]{0.48\columnwidth}
    \centering
    \includegraphics[width=0.9\textwidth]
    {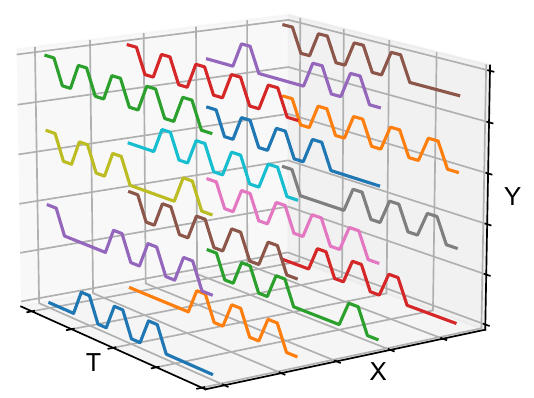}
    \caption{}
     \end{subfigure}
     \hfill
     \begin{subfigure}[b]{0.48\columnwidth}
    \centering
    \includegraphics[width=0.9\textwidth]{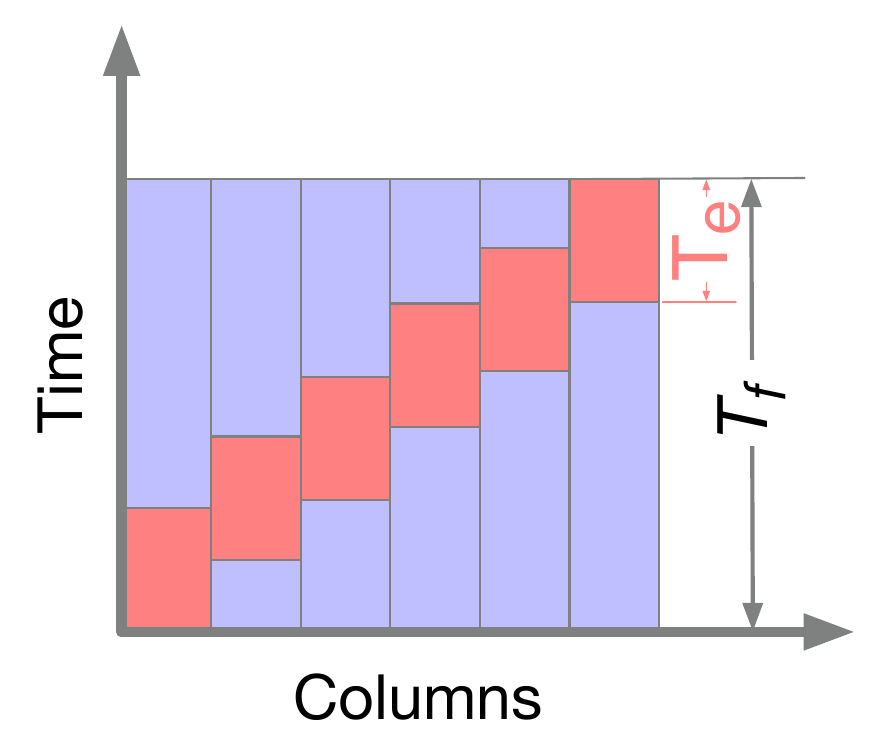}
     \caption{}
     \end{subfigure} 
     \caption{(a) The pixels in event cameras can independently and continuously capture changes in light intensity, where $X,Y$ are the pixel positions and $T$ is the timestamp. (b) During the exposure time $T_{e}$, only one column of pixels is sampled based on the rolling shutter mechanism of COMS cameras, while $T_{f}$ represents the frame rate.}
     \label{fig:comparision}
\end{figure}
Cameras are a popular choice for supporting multiple channels due to their high spatial resolution. Conventional cameras based on Complementary Metal-Oxide-Semiconductor (CMOS) or Charge-Coupled Devices (CCD) can essentially be viewed as arrays of photo-diodes, capable of measuring light information in different locations or pixels simultaneously. These cameras are typically designed for human vision applications like video recording, and their frame rates commonly range from 24 to 60 frames per second (fps)~\cite{poynton2012digital}. Although the rolling shutter mechanism, \yx{as shown in Fig.~\ref{fig:comparision} (b),} can enhance the sampling rate up to tens of kHz~\cite{yang2017ceilingtalk}, it restricts the target view to a specific pixel range, \yx{usually a column or a row}, thereby losing information from other pixels. Instead of using rolling shutter effect~\cite{hamagami2021rolling,kamiya2022visible,lee2015rollinglight,kuo2014luxapose}, there are also specialized scientific cameras can achieve higher frame rates up to 2k fps~\cite{iwase2014improving}. 
However, high-speed cameras capture the absolute status of every pixel in each frame, generating redundant data that increase computational complexity, storage requirements and \yx{transmission bandwidth burdens}.

Recent advancements in dynamic vision sensor have facilitated the measurement of light information in individual pixels based on changes in their state across subsequent frames. This development allows for significantly higher frame rates (up to approximately 1MHz) and lower energy consumption (10 mW) compared to traditional CMOS and CCD cameras~\cite{gallego2020event}. Considering these advantages, we designed our passive VLC receivers based on the dynamic vision sensor, also known as an event camera ~\cite{rebecq2019high}. 

Each pixel in an event camera~\cite{gallego2020event} operates independently. Here, each pixel continuously monitors the relative light intensity values and generates an \textit{event} once the intensity change exceeds a predefined threshold. Each event contains four numbers (\textit{timestamp, x, y, polarity}), where timestamp records when the change happens, $x$ and $y$ denote the pixel's 2D position, and polarity indicates the direction of the light change, with ON representing a brighter event and OFF representing a darker event.
Event cameras asynchronously generate events triggered by changes in brightness at individual pixels, without the limitations of frame rate, \yx{as shown in Fig.~\ref{fig:comparision} (a)}. This approach discards redundant information across consecutive frames and reduces power consumption by only processing the changing pixels, rather than the entire field of view of the image sensor. There are also other advantages associated with event cameras, such as low latency since there is no global exposure limit; as soon as changes occur, events are emitted. Additionally, event cameras offer high temporal resolution, with the minimum time difference between two events being as low as 1 $\mu$s.

\subsection{Duplicate Events}
\label{subsubsec:eventCameraPrinciples}
\begin{figure}[htp!]
    \centering
    \includegraphics[width=0.75\columnwidth]{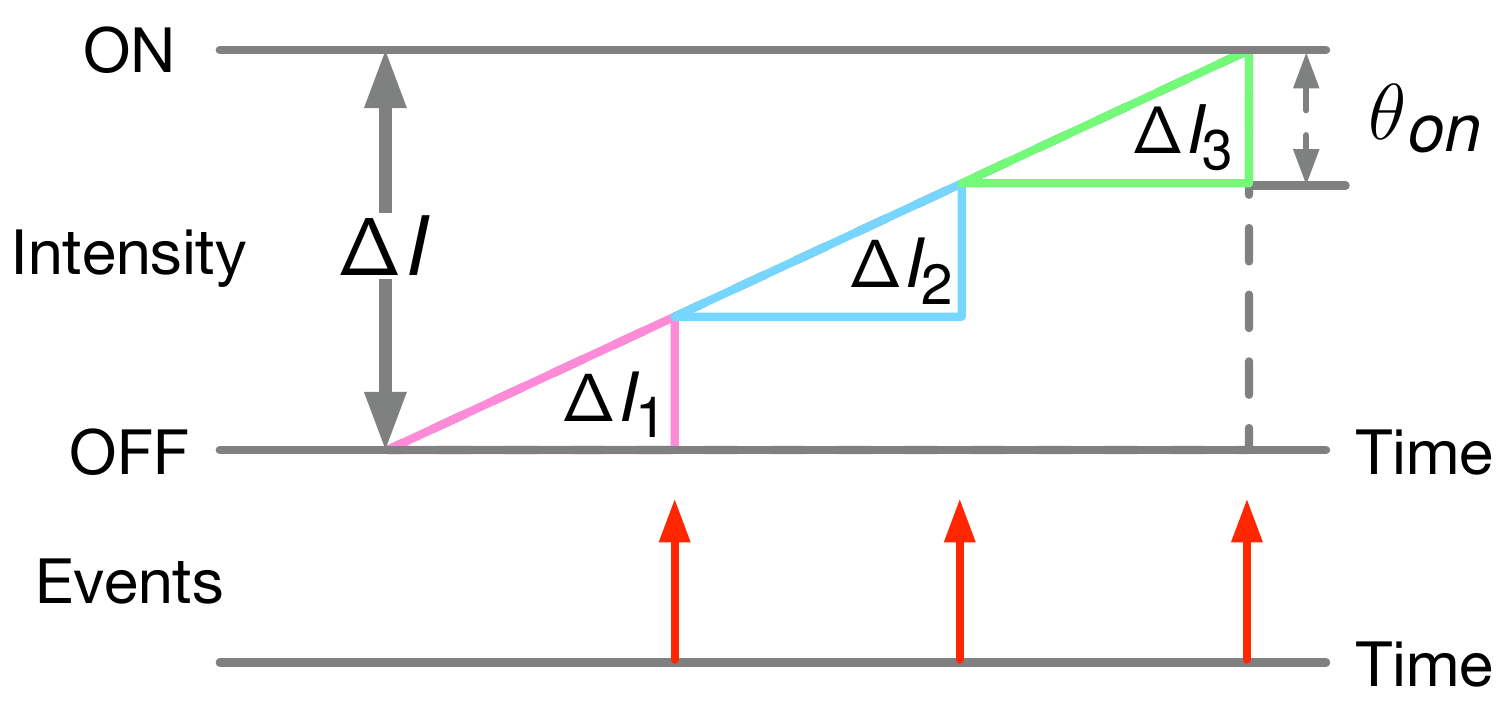}
    \caption{The large intensity change may trigger multiple events to be generated.}
    \label{fig:slow-process}
\end{figure}

\begin{figure}
    \centering
    \includegraphics[width = 0.55\columnwidth]{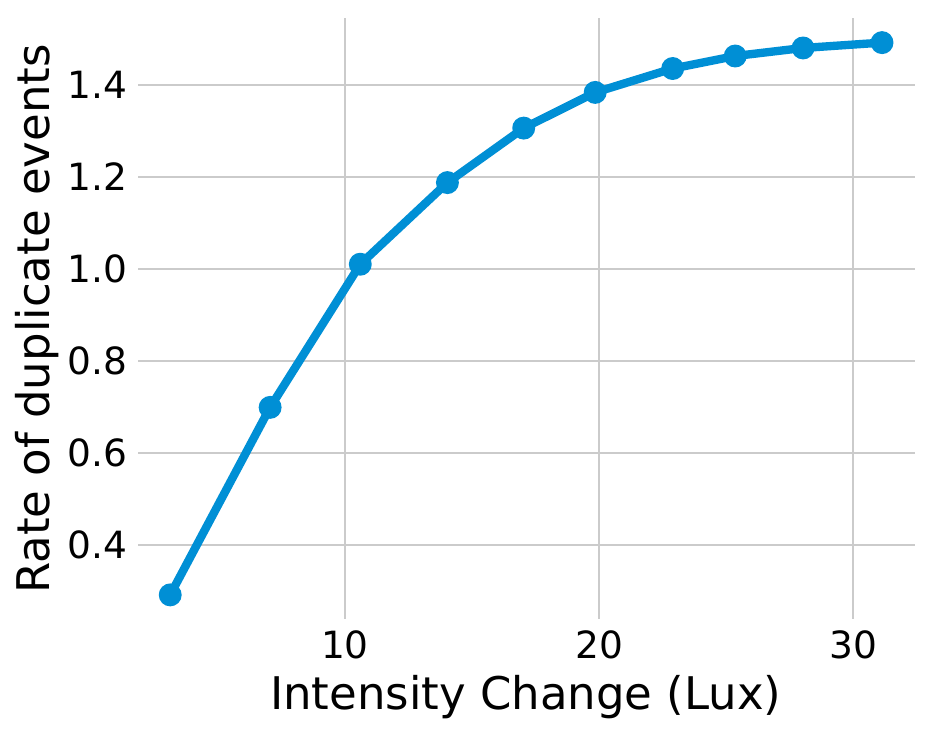}
    \caption{The ratio between duplicated events and unique events vs. the received
    light intensity change. }
    \label{fig:ratio-duplicate}
\end{figure}

As discussed above that the pixel in the event camera monitors the intensity changes, and it emits the event once the change overpasses a predefined threshold. Here, we dive deeper into the physical circuit to understand the relationship between intensity changes and event generations, which provides insights into the bandwidth limitation of the proposed VLC system. The working pipeline of one pixel can be abstracted into three steps as follows. First, the photo-detector converts the light signal into a photo-current $I_{p}$ along with a dark current: $I=I_{p}+I_{dark}$, in which the $I_{dark}$ is proportional to the power of light signal. The current $I$ is then converted to a voltage $V_p$ logarithmically. Secondly, $V_p$ is amplified and compared with the previously memorized value. Thirdly, when the difference $\Delta V_p$ exceeds the threshold $\theta_{on}$ or $\theta_{off}$, an event is generated. This process can be expressed mathematically as:
\begin{align}
    \label{eq:delta_intensity}
    &  \Delta V_{x,y} = log(I(x,y)_{t2})-log(I(x,y)_{t1})
    \\
    &  \left\{
\begin{array}{l}
  ON\,event\; \quad When \quad \Delta V_{x,y} \geq  \theta_{on}, \\
  NO\,event\; \quad When \quad -\theta_{off} < \Delta V_{x,y} < \theta_{on}, \\
  OFF\,event \quad When \quad \Delta V_{x,y} \leqslant  -\theta_{off},
\end{array}
\right.
\end{align}
where $x,y$ represents the pixel position, $t_1$ and $t_2$ are two timestamps, and $t_1 < t_2$. 
We note that when the intensity change $\Delta I$ is large enough, multiple monotonic events will be triggered. The reason is that $\Delta I$ can be decomposed into multiple small $\Delta i$ that cause event generation. In other words, the number of generated events is proportional to $\Delta I$. For instance, in Fig.~\ref{fig:slow-process}, the intensity change $\Delta I$ comprises three minor processes whose changes exceed the threshold $\theta_{on}$, which are subsequently transformed into three discrete events. For negative intensity changes, the situation is similar. \textbf{Therefore, beyond the initial event triggered by each change, subsequent repeated events may also occur for the same lighting change.}
Our experiment result (Fig.~\ref{fig:ratio-duplicate}) shows that the number of duplicate events can be more than that of unique events when 
the light intensity (received signal strength) change ($\Delta$) at the receiver is high (i.e., more than 10 Lux).

\emph{Observation 1: This unique phenomenon of duplicate signals does not occur in prior VLC systems, where pixels in CMOS and CCD cameras, or photo diodes, are sampled only once per frame or time slot, respectively.} This needs to be considered in the decoding process, which will be introduced later in Section~\ref{sec:duplicate}.

\subsection{Delayed Timestamping of Events}
\label{sec:delay}

After events are generated by the event camera's pixels in response to brightness changes, adding time stamps and being read out are critical processes. Event cameras employ a novel asynchronous event readout scheme~\cite{inivation2020understanding} that differs fundamentally from their traditional synchronous counterparts.
Synchronous readout methods, used in conventional frame-based cameras, are characterized by a fixed maximum delay inversely proportional to the sampling rate. However, this approach is prone to generating redundant data during periods of minimal or no activity in the scene. Even when there are no brightness changes occurring, the camera continues capturing entire frames at regular intervals, resulting in a significant amount of redundant information.
In contrast, the asynchronous readout scheme implemented in event cameras operates in an event-driven manner. It selectively records only the pixels where the intensity change exceeds a predefined threshold. Events at pixels with insignificant brightness variations are ignored. This approach significantly reduces data temporal redundancy.

The process of adding time stamps for events in this asynchronous context is governed by two key mechanisms: event detection and readout~\cite{inivation2020understanding}. A critical aspect to consider in the event detection is the inherent variability due to physical circuit constraints. Specifically, events simultaneously triggered by the same stimulus can exhibit non-uniform timestamp differences across various pixels, indicating different priorities in the event generation and readout process.
Additionally, after the events are detected or generated, they will be queued in the chip and then processed by the micro-controller unit (MCU) to add time stamps. It should be noted that the time stamps are not the ``exact'' time when the events are produced but the time they are read out by the MCU~\cite{inivation2020understanding}. This readout delay is intimately tied to the throughput/bandwidth capacity of the specific camera in use. Together, the detection delay and readout delay consist of the delay of the event time stamps.

\emph{Observation 2: The nondeterministic delay in event timestamping poses unique challenges for receiver synchronization during demodulation, particularly in high data rate scenarios, where the delay spread is more acute. This issue must be addressed when designing demodulation for event camera-based VLC receivers.}
\begin{figure}[htp!]
    
    \begin{subfigure}[b]{0.49\columnwidth}
    \centering
    \includegraphics[width=\textwidth]{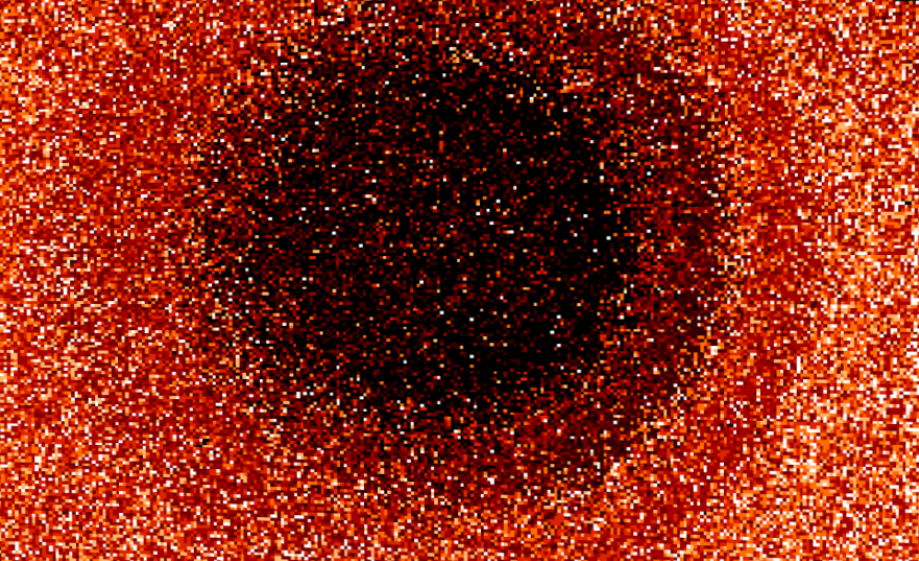}
    \caption{ON events. }
    \end{subfigure}
    \hfill
    \begin{subfigure}[b]{0.49\columnwidth}
    \centering
    \includegraphics[width=\textwidth]{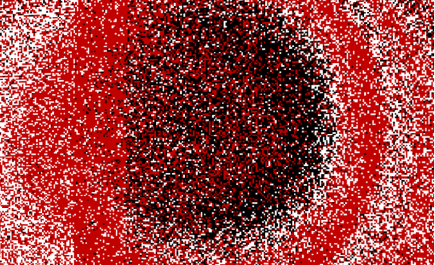}
    \caption{OFF events. }
    \end{subfigure}
    \caption{Spatial distribution of event timestamping delays, where the color indicates the normalized delay; black represents zero delay and white denotes maximum delay.}
    \label{fig:slice}
\end{figure}

\subsection{Center-first Pixel Readout Priority}
\label{sec:center_peripheral}
From Section~\ref{sec:delay}, we understand that timestamps of events triggered by the same stimulus at different pixels are not identical. To further explore the distribution of these differences across the spatial domain, we activate all mirrors in the DMD to initiate a flash, maintaining this state for one second to allow sufficient time for reading out all events. By treating the moment of state change as the reference or starting point, we can calculate the delays of the produced events. These delays are depicted in Fig.~\ref{fig:slice}, where the color of each pixel represents the duration of the delay. As illustrated in the figure, in contrast to conventional rolling shutter cameras that scan regions line-by-line, the central circular region of the event camera demonstrates a shorter delay, as indicated by colors closer to black, while the peripheral areas exhibit longer delays, with colors approaching white. This phenomenon is observed for both ON and OFF events. The slight difference in the spatial delay patterns between ON and OFF events can be attributed to their distinct threshold values. \yx{Furthermore, similar effects occur when utilizing lenses with varying focal lengths and apertures.}

\emph{Observation 3: This distinctive pixel readout behavior of event cameras presents an opportunity for us to encode data at varying refresh rates across different regions of the DMD transmitters, thereby optimizing overall data transmission rates.} 

\subsection{Event Loss}
\label{subsec:eventLoss}

\begin{figure}[htp!]
    \centering
    \includegraphics[width=\columnwidth]{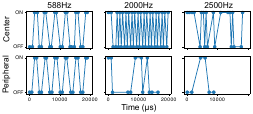}
    \caption{Events at center and peripheral channels at different refresh rates.}
    \label{fig:low-vs-high}
\end{figure}

Having established that the event camera prioritizes the central region over the peripheral when generating the timestamping the events, we further investigate the impact of timestamping delay on potential event loss when the light is frequently turned ON and OFF. To this end, we configure the DMD to operate at both low (588 Hz) and high refresh rates (2,000 Hz and 2,500 Hz), continuously alternating the state of the mirrors. We select two pixels for analysis: one located near the optical center (the geometric center of the camera's pixel array) and another situated near the pixel array's edge. The events generated by these pixels are illustrated in Fig.~\ref{fig:low-vs-high}.

At low refresh rates, event capture is efficient across both central and peripheral channels. However, as the refresh rate increases, event loss begins to occur, with the peripheral region suffering more significantly than the central area. This disparity can be attributed to the higher refresh rates generating more events per unit time, leading to increased event congestion. This congestion, in turn, amplifies timestamping delays for more events in the periphery than the center. Events delayed beyond the next refresh cycle are inevitably lost, causing more loss in the periphery. To explore whether exposure time influences event loss, we conducted an experiment at a refresh rate of 2,500 Hz, equating to an ON state duration of 400 $\mu$s, and extended the OFF period between ON states to one second. Surprisingly, both ON and OFF events were accurately captured at their respective frequencies indicating that event loss does not stem from insufficient exposure time.

\emph{Observation 4: DMD refresh rates would be constrained by the readout speed or the bandwidth of the event cameras. Operating the DMD beyond a threshold refresh rate can lead to event loss and eventually bit errors as the decoding is dependent on event information such as time stamps and event types (ON or OFF).}

\section{system design}
\label{sec:system}
\subsection{System architecture}
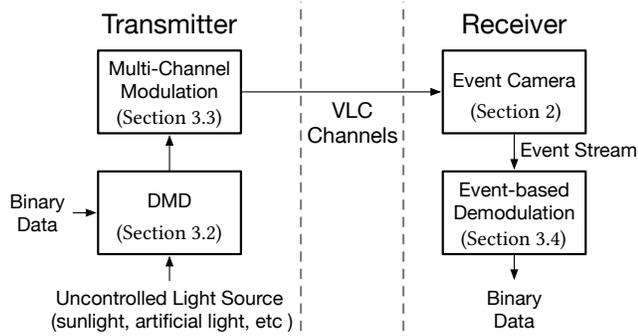
\begin{figure}[htp!]
    \centering
    \begin{overpic}[width=\columnwidth]{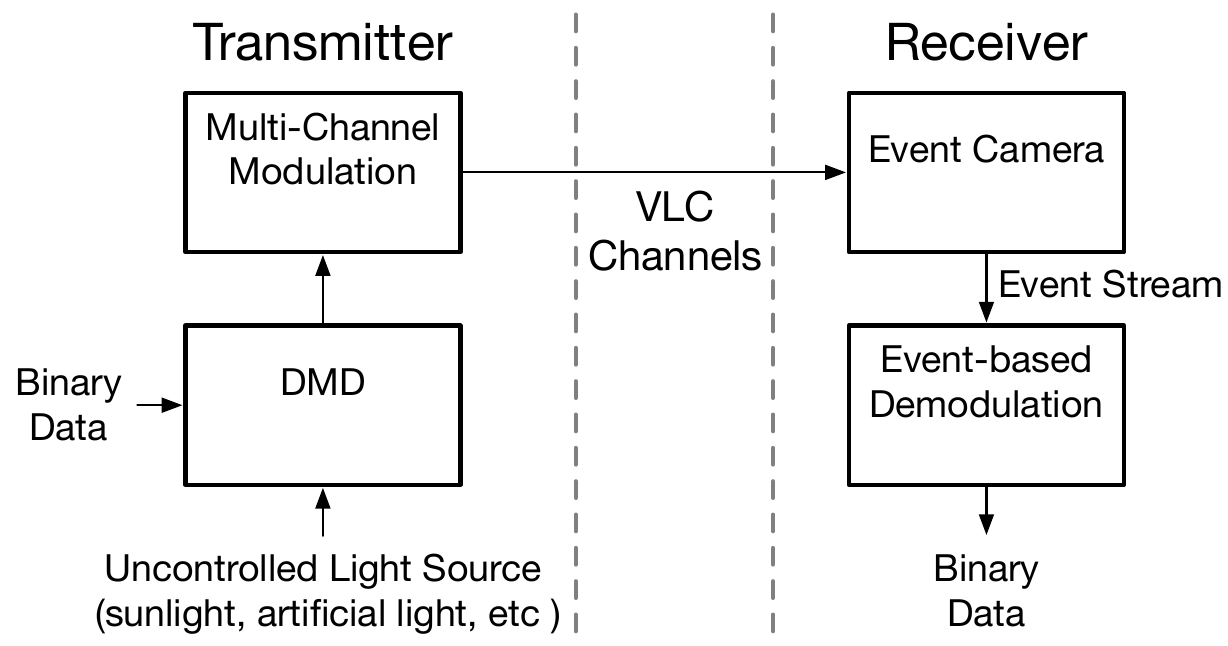}
    \put(18,15.5){\small (Section \ref{sec:transmitter})}
    \put(18,33.5){\small (Section \ref{sec:multi-c})}
    \put(73.5,34.5){\small (Section \ref{sec:receiver})}
    \put(72,14.5){\small (Section \ref{sec:demodulation})}
    \end{overpic}
    \caption{System overview of \cname.}
    \label{fig:system}
\end{figure}
The \cname system shares the same foundational components as other passive VLC systems, including a uncontrolled light source, a transmitter, and a receiver (see Fig.~\ref{fig:system}). 
\yx{The incident light from the light source} is directed towards the micro-mirror array within the DMD chips. Through multi-channel modulation, the binary data is transmitted via multiple channels, i.e., spatially separated blocks of designated micro-mirrors. Subsequently, the event camera captures the light signals reflected by the mirror blocks and outputs event streams. A novel event-based demodulation is then performed on the events to reconstruct the binary data carried by the events.
Our novel contributions are centered around three main objectives: (1) configuring the mirrors into a large number of non-interfering blocks or channels to achieve high data rates, (2) mapping DMD mirror blocks or channels to unique block of event camera pixels, and (3) decoding the binary data directly from the event stream generated by the event camera. 

\subsection{Transmitter}
\label{sec:transmitter}
\begin{figure}[htp!]
    \centering
    \includegraphics[width=0.7\columnwidth]{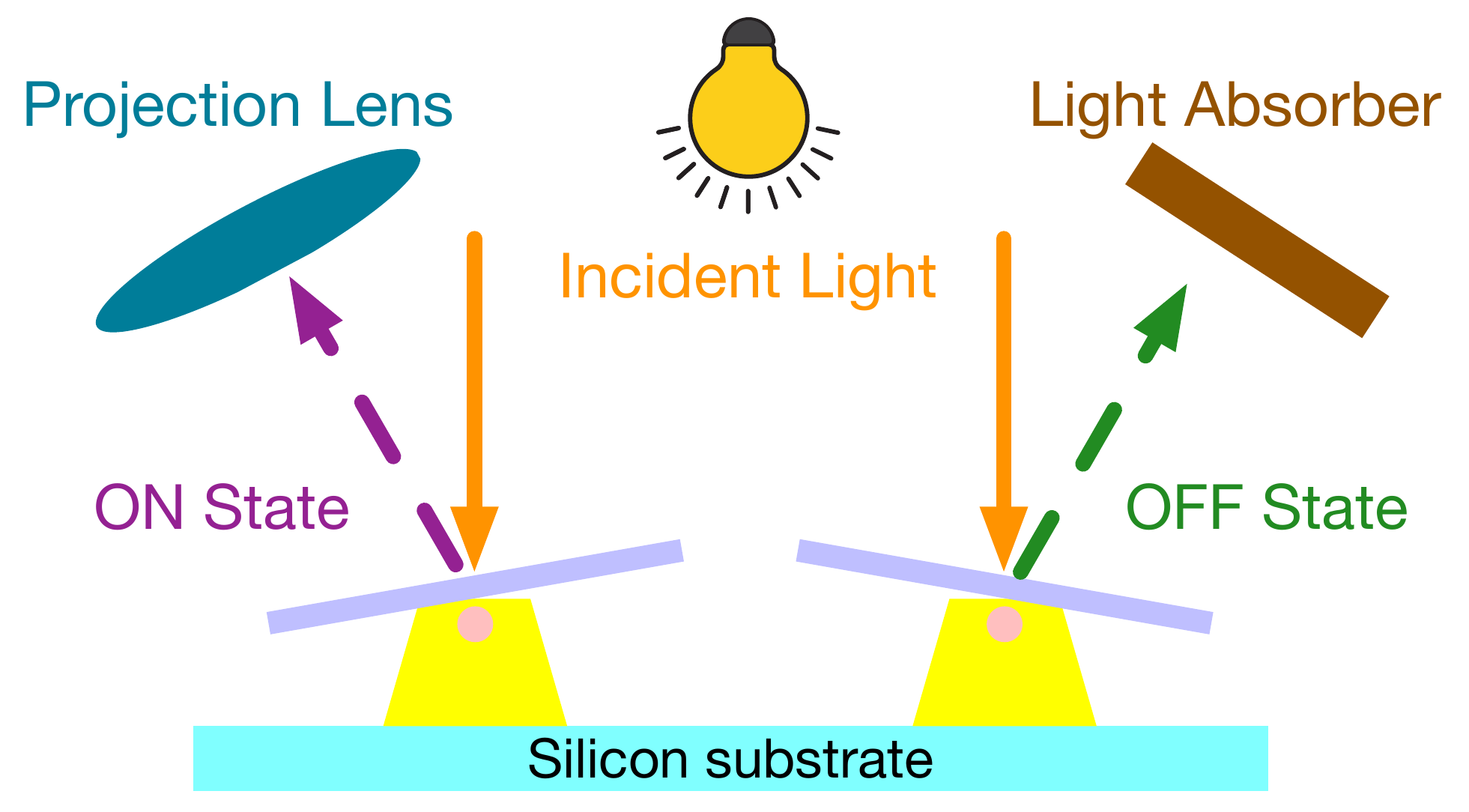}
    \caption{ON/OFF states of a micro-mirror in a DMD.}
    \label{fig:dmd}
\end{figure}
\cname employs a DMD as the transmitter, an optical system consisting of millions of micro mirrors arranged in a rectangular array. Every mirror can be tilted to two states: $\pm 12^{\circ}$, i.e., either ON or OFF. When in the ON state, a mirror redirects the incident light towards the projection (out) lens. Conversely, the light is absorbed by the internal materials of the DMD when a mirror is in an OFF state, as shown in Fig~\ref{fig:dmd}. DMDs have been extensively utilized in digital projections~\cite{pan2007homogenized}, single-pixel imaging~\cite{duarte2008single}, and structural light applications~\cite{li2008accurate}. This widespread adoption as a commodity has significantly reduced their cost, often priced at approximately \$40 each in bulk~\cite{xu2022exploiting}. Different DMD chips have different numbers of mirrors and maximum fresh rate. Compared to LCD shutters, the attenuation of light propagation is lower (3\% for DMD and 50\% for LCD shutters)~\cite{dmd,ghiasi2021principled}. Also, since DMD controls the reflection surface, there are no limitations for the light source, which means any kind of light can be used or modulated, e.g., sunlight, laser light and artificial light. 
To control the micro-mirrors, the DMD chip reads binary data that dictates the state of each corresponding mirror: `1' for the ON state and `0' for the OFF state. The high switching speed and precise control of individual micro-mirrors enable DMDs to generate intricate light patterns.

\subsection{Multi-Channel Modulation}
\label{sec:multi-c}
\subsubsection{OOK for Multiple Channels}
\label{sec:OOK}
\begin{figure}[htp!]
    \centering
    \includegraphics[width=0.65\columnwidth]{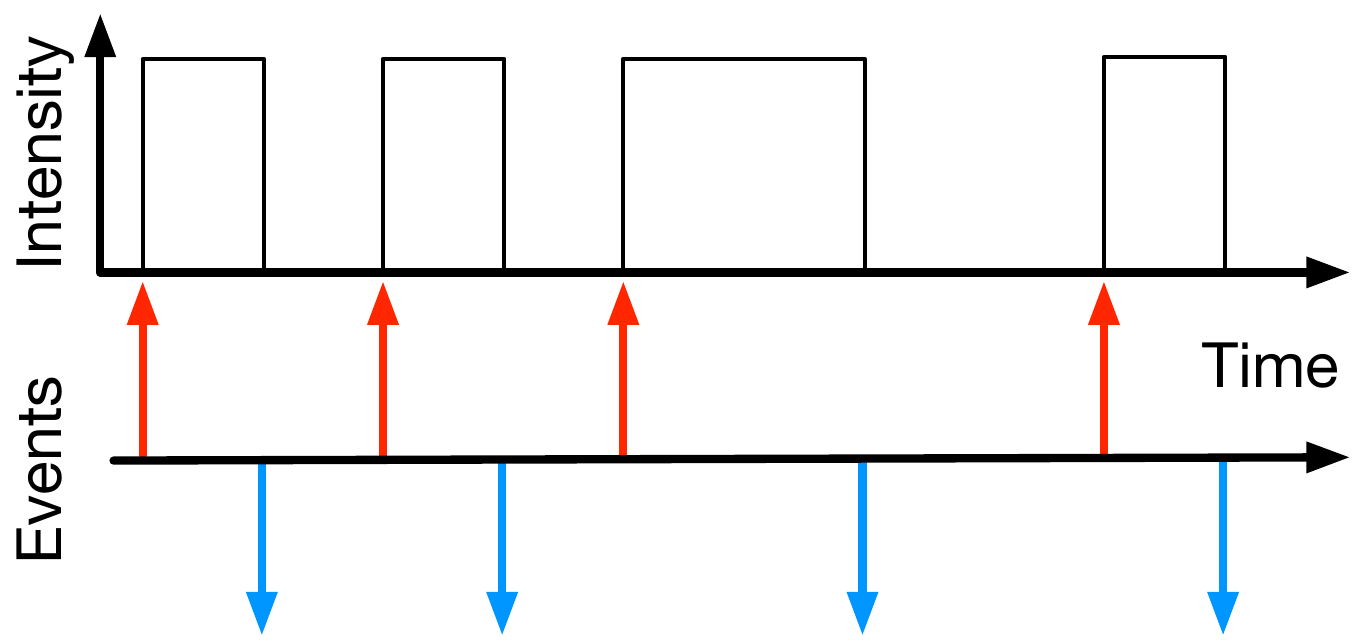}
    \caption{The illustrations on relationships between intensity changes and the corresponding events. The first row is the intensity value, while the second row is the event stream. The red arrows are ``ON" events and the blue ones are ``OFF" events. }
    \label{fig:ook-decode}
\end{figure}
In the context of VLC, 
as elaborated above, 
both DMD and event cameras demonstrate a heightened sensitivity to transitions between two distinct states. This characteristic renders the On-Off Keying (OOK) modulation method particularly suitable for such systems. 

In theory, each micro-mirror within a DMD can function as an independent transmitter. However, a practical limitation arises due to the relatively weak light output of a single mirror, which is insufficient for reliable detection by the receiver. To overcome this, we grouped the DMD mirrors into larger blocks as shown in Fig.~\ref{fig:mchannel}, each block consisting of a predefined number of mirrors, for instance, a $ M \times M$ array, where $M$ is the number of mirrors in a row and column. This approach leverages these blocks as independent channels, enabling the parallel transmission of distinct bit streams. Within each block, the mirrors are manipulated to represent binary states: the ON and OFF states correspond to transmitting a bit `1' and `0', respectively. %
Additionally, it is crucial to note that event generation is triggered specifically by the rising and falling edges of the signal, as illustrated in Fig.~\ref{fig:ook-decode}. This differs from conventional OOK schemes, which rely on absolute amplitude values at the receiver to decode data.
Therefore, the overall data rate of this system can be quantified by :
\begin{equation}
\label{eq:datarate}
\text{Data Rate} = N \times Y \times \log_{2}X,
\end{equation}

where $N$ represents the number of channels, $Y$ denotes the refresh rate of the DMD (measured as the baud rate), and $X$ indicates the number of distinct symbols used in modulation. For the OOK scheme, where there are two symbols, i.e., ON and OFF, $X = 2$. This formula highlights that an increase in the number of channels ($N$) and refresh rate ($Y$) can linearly augment the system's throughput. 

Moreover, this method can be seen as a practical implementation of the multiple-input and multiple-output (MIMO) concept, specifically through the technique of multiplexing. To mitigate interference between adjacent channels, a security margin, also known as a guard gap, is implemented. In these margin areas, the mirrors are maintained in the `off' state to prevent signal overlap.

To achieve higher data rates, it is advantageous to maximize the value of $N$. However, the potential for increasing $N$ is inherently limited by the resolution of the transmitter. In essence, enhancing $N$ necessitates a reduction in $M$, the individual mirror count in a row per block, which unfortunately leads to weaker signal strength. An alternative approach to augmenting data rates is to increase $Y$, the refresh rate. However, this solution is not without its drawbacks, as it results in shorter exposure time and a higher event count, which can consequently introduce longer delays and event loss (refer to Observation 4 in Section~\ref{sec:receiver}). 

\subsubsection{Single vs Dual Refresh Rate Configurations}
\label{sec:dual}
\begin{figure}[htp!]
\centering
\includegraphics[width=0.7\columnwidth]{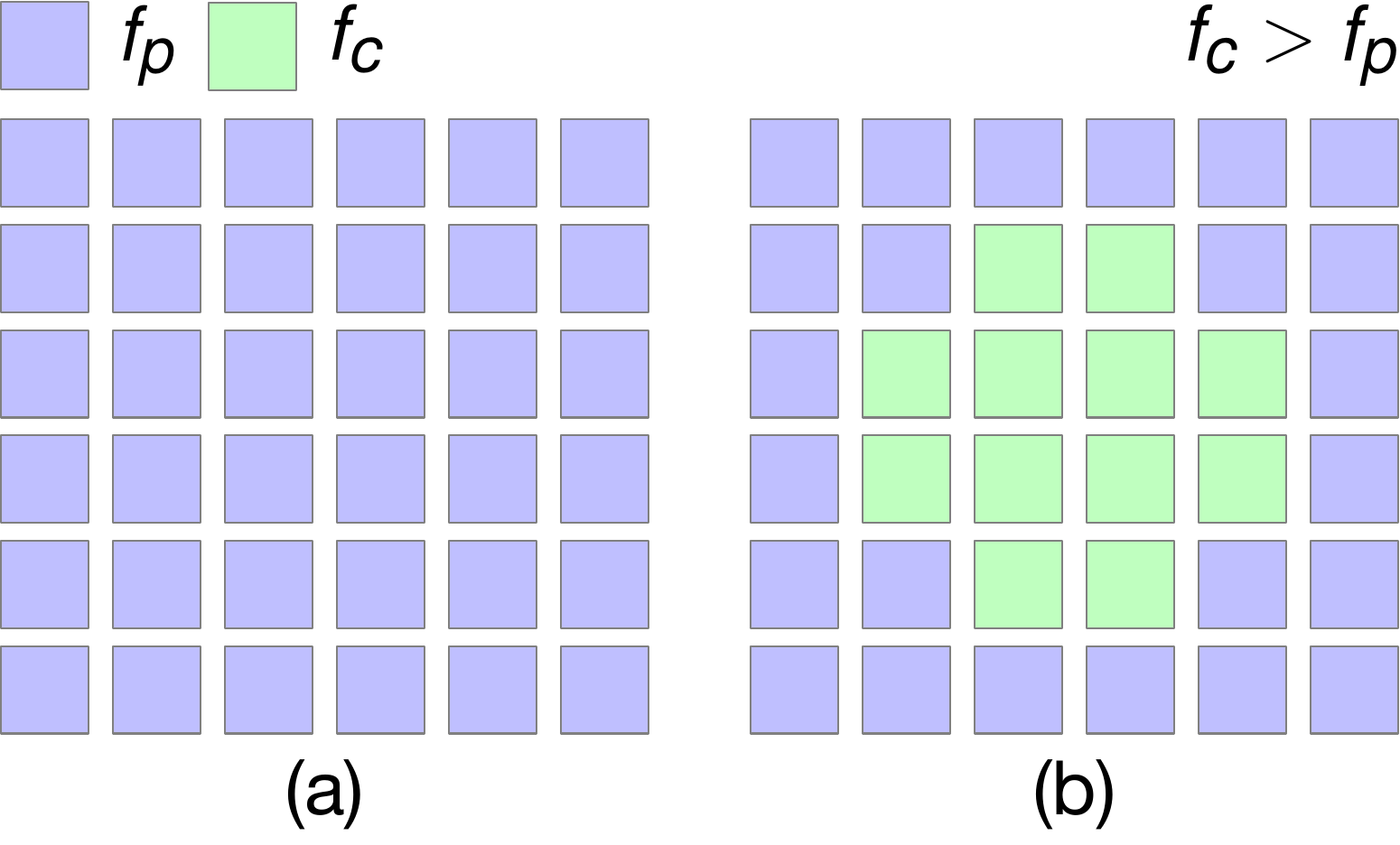}
\caption{Single (a) and dual (b) refresh rate configurations.}
\label{fig:diff-fre}
\end{figure}

As discussed in Section~\ref{sec:OOK} earlier, the overall data rate of an \cname system is a function of the number of channels and their respective refresh rates. To optimize for maximum throughput, the configuration of refresh rates is crucial. A na\"ive method is to set the channels to operate uniformly at a single frequency, $f_{p}$, as depicted in Fig.~\ref{fig:diff-fre} (a), with each block representing one channel. However, our insights in Section~\ref{sec:center_peripheral} (Observation 3) suggest that the central area of the receiver can respond more rapidly to changes. Maintaining a uniform refresh rate across all channels, including those in the center, fails to capitalize on their heightened temporal sensitivity. To harness the full potential of the receiver, we propose allowing channels to operate at dual refresh rates, $f_{p}$ and $f_{c}$, where $f_{c} > f_{p}$. 
Inspired by Fig.~\ref{fig:slice}, we employ a geometric constraint to select the appropriate centered channels at $f_{c}$. Specifically, we consider the central channel as the origin, denoted by $C_{0,0}$. A channel $C_{x,y}$ is selected if its Euclidean distance $R=\sqrt{x^{2}+y^{2}}$ from the origin is less than a predefined threshold $d$. Channels within this distance will operate at a higher refresh rate ($f_{c}$), while those beyond it will operate at a lower refresh rate ($f_{p}$). By adjusting the threshold $d$, we can manipulate the ratio of the number of channels operating at $f_{c}$ to those at $f_{p}$ to achieve the maximum overall system data rates.

\subsection{Event Based Demodulation}
\label{sec:demodulation}
The demodulation of \cname consists of three phases:

\ding{202}\textit{Channel Mapping}: This initial phase utilizes predefined data paradigms. During calibration, the receiver extracts crucial parameters needed for demodulation, specifically, the channels' positions and channel indexes.
    
\ding{203}\textit{Duplicate event removal}: When the intensity change exceeds the change threshold, multiple events will be generated in response to a single stimulus as discussed in Fig.~\ref{fig:ratio-duplicate}, Section~\ref{sec:receiver}. However, since we are primarily interested in the time at which the state transitions occur, which is indicated by the first event changing polarity, the subsequent duplicated events need to be filtered out as they do not provide additional information.

\ding{204}\textit{Decoding}: In this phase, event streams are segmented into different channels before being decoded. 
    Here, we propose two decoding algorithms that use the timestamps in the events differently.

\subsubsection{Channel Mapping}

To establish multi-channel communication, the initial step involves determining the channel mapping relationship between the transmitter and the receiver. 

The incident light from sources reflects off DMD micro-mirrors, projecting onto a plane. An event camera captures this projection. However, size differences and potential misalignment between the projection plane and camera's pixel array result in unknown channel distribution at the receiver end.
Traditionally, camera calibration is performed to address this issue, which involves complex 3D geometric algorithms and the use of an auxiliary tool, such as a chessboard, as referenced in~\cite{muglikar2021calibrate,huang2021dynamic}. In contrast, we propose a data-driven channel mapping approach that allows for rapid extraction of the channel mapping (less than two seconds) without the need for additional tools.
\begin{figure}[htp!]
    \begin{minipage}[t]{0.48\columnwidth}
    \centering
    \includegraphics[width=0.7\textwidth]{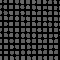}
    \caption{The heat map of event counts.}
    \label{fig:heatmap}
     \end{minipage}
     \hfill
     \begin{minipage}[t]{0.48\columnwidth}
    \centering
    \includegraphics[width=0.7\textwidth]{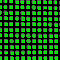}
    \caption{The bounding box of founded channels.}
    \label{fig:boudning}
     \end{minipage}
\end{figure}

We initiate the process by transmitting distinct packets across all channels at a low refresh rate (e.g., 588 Hz) for one second, with each packet's payload being the channel number. This transmission results in the generation of events for the pixels corresponding to active channels. By tallying the number of events for each pixel, we create a heat map, as depicted in Fig.~\ref{fig:heatmap}. This map uses the event counts to represent pixel intensity, allowing us to discern clusters of pixels associated with the same channels. Pixels that appear black on the heat map indicate the absence of channels and, consequently, no event generation (the blank pixels in Fig.~\ref{fig:heatmap}). Subsequently, we pinpoint the exact locations of the channels by employing bounding box extraction algorithms~\cite{suzuki1985topological}. This process involves a series of image processing techniques, including binarization, erosion, and dilation, using the OpenCV library\cite{bradski2000opencv}. As illustrated in Fig.~\ref{fig:boudning}, our approach successfully and precisely identifies the positions of the channels.

Given that we have already identified the area corresponding to each channel, we select its geometric center as the target point for decoding, or the pixel of interest. Once the payload is decoded, we can ascertain the channel index. This method, being based on data, imposes no stringent requirements on the cameras or DMD and is adaptable to any environment. The mapping between channels and their positions remains constant unless there is a change in their relative positions. Should such a change occur, the aforementioned process should be repeated.

\subsubsection{Duplicate Event Removal}
\label{sec:duplicate}
As discussed in Section 
\ref{subsubsec:eventCameraPrinciples} (Observation 1), duplicate events will be generated when the intensity change is significant. These duplicated events should be filtered out during the decoding phase, as they do not carry any additional information.
Specifically, consider an event stream [$e_{1}$, $e_{2}$, ..., $e_{n}$] that occurs at Pixel $(x,y)$, where $e_{i}$ represents the event that occurred at time $i$. Only events where the polarity of $e_{i}$ differs from $e_{i-1}$ will be retained; all others will be filtered out. The first event should always be included at the beginning of the stream.
After this step, the ON OFF events will happen in turns in one event stream as there is no repeated events with the same polarity.

\begin{figure}[htp!]
    \centering
    \includegraphics[width=\columnwidth]{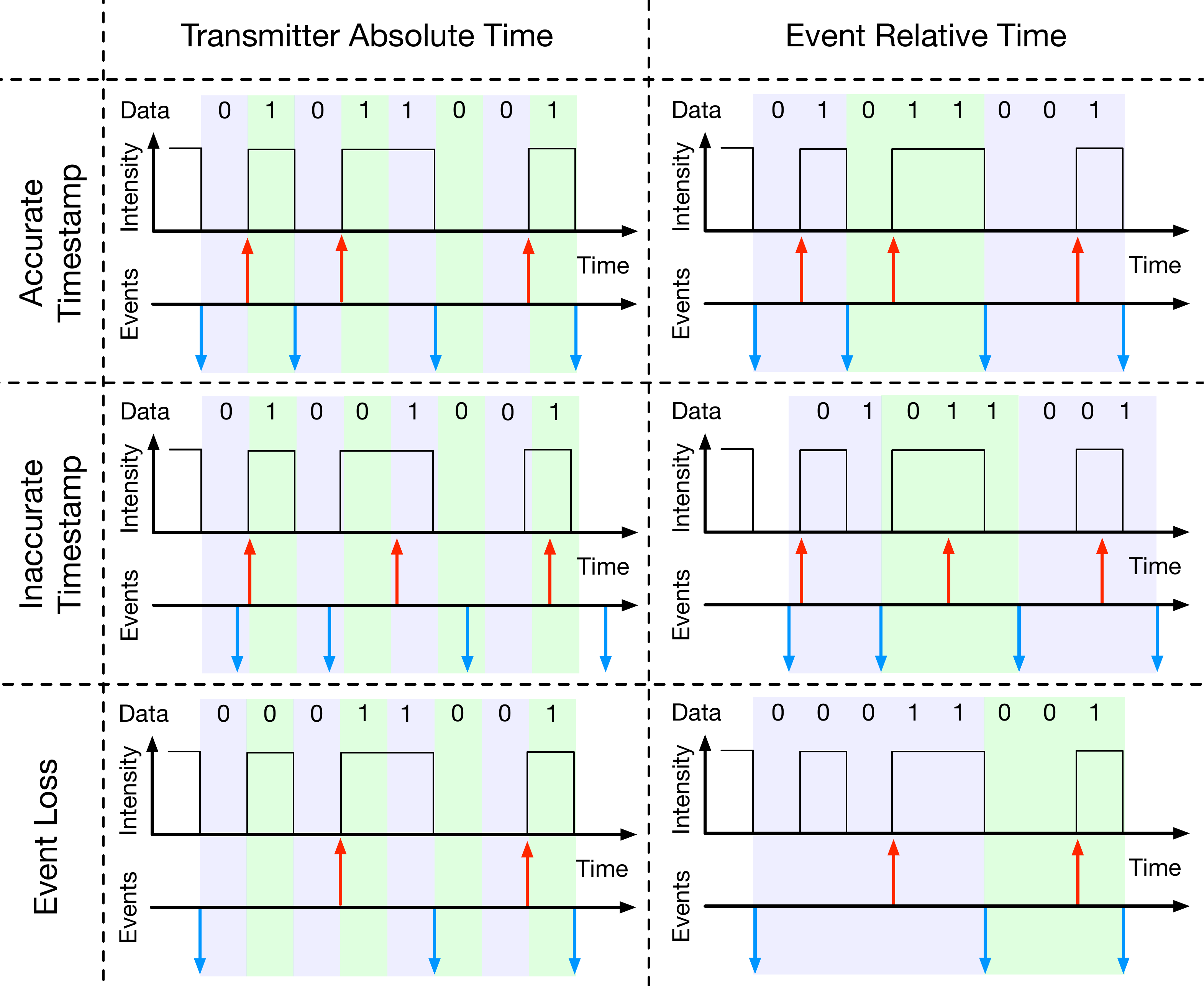}
    \caption{Decoding with the transmitter absolute time and the event relative relative time, with accurate, inaccurate timestamps and event loss.}
    \label{fig:global}
\end{figure}

\subsubsection{Decoding}
\label{subsubsec:decoding}
After the channel mapping step, the channels' positions are determined. Next, we decode the binary data from received event streams. %
Recall that each channel consists of multiple micro mirrors, and these mirrors transmit the same data. So, we may choose any of pixel in a channel to decode the data. However, the events in different pixels 
in the same channels may not be the same due to noise and interference. The pixel in the geometric center of the channel is thus a good candidate since it is the farthest from the neighboring channels and the interference is the smallest. 
Furthermore, the light in one channel will aggregate, and the center receiving point has the maximum gain. Based on this observation, we perform decoding based on the events at the center pixel of the channel only.
Then, the four tuple data $(t,x,y,p)$ can be simplified to two dimensions $(t,p)$, in which $t$ and $p$ are the timestamp and polarity (event type, ON or OFF), respectively.

Next, we propose a decoding algorithm based on these ``events'' data. 
Different from the conventional OOK demodulation that utilizes the intensity amplitudes to determine the bit value, we will decode the data based on the \textbf{timestamps} of the events. Specifically, we design two different timestamp event decoding methods as follows.

\textbf{Transmitter Absolute Time (Baseline).} This method follows the conventional demodulation pipeline. The first step is to synchronize the receiver's clock with the transmitter's clock by exploiting the preamble. Then, we apply a sliding time window. 
The window duration is approximately the symbol duration, which is $\frac{1}{Y}$ as shown in Fig.~\ref{fig:global}.
Let $b_i$ denote the $i$-th decoded bit, $t_i$ represent the time of the $i$-th event, and $T_s = \frac{1}{Y}$ be the symbol duration, where $Y$ is the baud rate.
The decoded bit $b_i$ can be expressed as:
\begin{equation*}
b_i = \begin{cases}
1, & \text{if an ON event is detected within } [t_i, t_i + T_s] \\
0, & \text{if an OFF event is detected within } [t_i, t_i + T_s] \\
b_{i-1}, & \text{if no event is detected within } [t_i, t_i + T_s]
\end{cases}
\end{equation*}
where $b_{i-1}$ represents the previous decoded bit. If there is no event detected in $b_0$, the last bit of preamble will be used accordingly. Fig.~\ref{fig:global} (Top Row, First Column) shows an 
example of transmitter absolute time decoding for \emph{0b01011001}.

\textbf{Event Relative Time.} Because event cameras read out events asynchronously, and the timestamp can experience additional read out delays in the MCU's processing queue, particularly during high data rates, as discussed in Section~\ref{sec:delay}, the event timestamp may become desynchronized from the transmitter's clock, even if they were initially synchronized during the preamble. Therefore, our second method explores the relative time difference between two OFF events. We determine the bit length by comparing the time difference between two OFF events rather than ON events because electronic noise can falsely trigger ON events~\cite{hu2021v2e}. Besides, we know there is always one ON event between two OFF events after \textbf{duplicate event removal}. 

Let $t_i^{\text{OFF}}$ denote the time of the $i$-th OFF event, $t^{\text{ON}}$ represent the time of the ON event between $i$-th and $i+1$-th OFF events, and $Y$ be the symbol rate.
The duration $T$ between two consecutive OFF events is given by:
\begin{equation}
T = t_{i+1}^{\text{OFF}} - t_i^{\text{OFF}}
\end{equation}
Then, the number of bits $H$ during the duration $T$ can be calculated as:
\begin{equation}
H =  \frac{T}{\frac{1}{Y}} = T \times Y.
\end{equation}
Let $N_1$, $N_0$ be the number of bits 1 and 0 within the duration $T$, respectively. These values can be determined by comparing the time differences\footnote{The floor function $\lfloor \cdot \rceil$ is used to round the result of the division to the nearest integer.} between ON and OFF events:
\begin{equation}
N_0 = \left\lfloor \frac{t^{\text{ON}} - t_i^{\text{OFF}}}{T} \times H \right\rceil,
\end{equation}
and
\begin{equation}
N_1 = H - N_0.
\end{equation}
The decoded data within the duration $T$ can be represented as a sequence of $N_0$ bits 0 followed by $N_1$ bits 1.
If no event is generated within the expected duration, the last decoded bit will be repeated. Fig.~\ref{fig:global} (Top Row, Second Column) shows an 
example of event relative time decoding for \emph{0b01011001}.

Moreover, we compare the two decoding methods under inaccurate timestamps (Observation 2 in Section~\ref{sec:delay}) and event loss (Observation 4 in Section~\ref{subsec:eventLoss}). For inaccurate timestamps caused by the asynchronous scheme (Fig.~\ref{fig:global} Second Row), the event relative time-based method can still extract the correct bits, although the time is randomly drifted, as far as the relative time difference is preserved. However, the transmitter absolute time approach relies on precise window slicing to determine the bits, which may suffer when more timestamps are inaccurate. Our results that will be introduced in Section~\ref{subsubsec:tatVsErt} later show that event relative time method can reduce the BER by 20 times when the data rate is high, i.e., 1.35 Mbps. Both approaches will face failure when events are lost as the transitions between states are missed (Fig.~\ref{fig:global} Third Row).

\section{evaluation}
\label{sec:evaluation}

\begin{figure}[htp!]
    \centering
    \includegraphics[width=0.8\columnwidth]{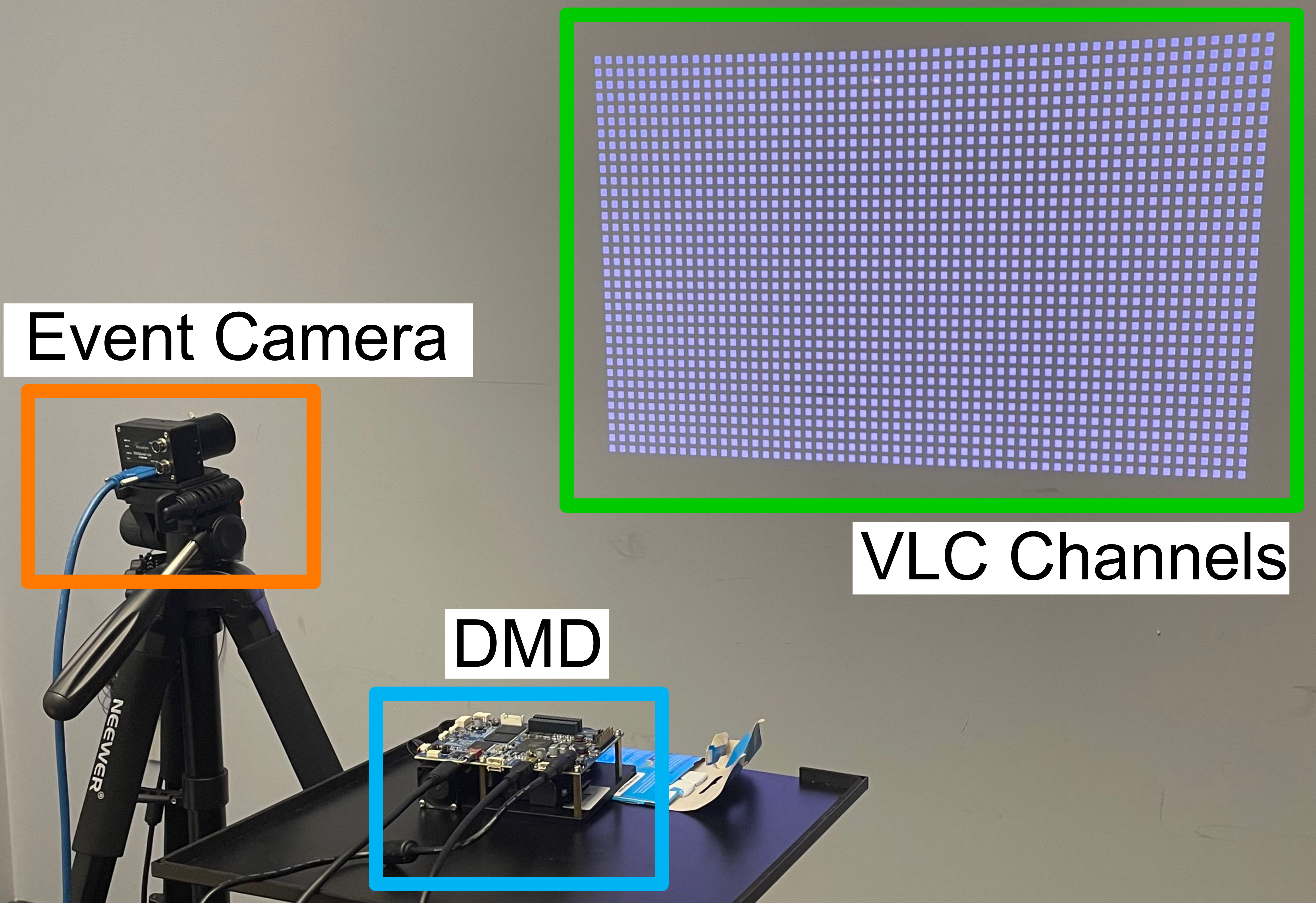}
    \caption{Experiment setup: Event camera reads multi-channel VLC data from DMD projection on a wall.}
    \label{fig:realsetup}
\end{figure}

\subsection{Prototyping and Experimental Setup} \textbf{Prototype:} \yx{Our prototype's transmitter is constructed with a DLP4500 Evaluation Module, which includes a DMD chip featuring 1,039,680 micro mirrors arranged in a 1,140 x 912 array, capable of refreshing at a maximum rate of 4,225 Hz.} \yx{Except for the DMD chip, there are also LED bulbs providing incident light and optical lenses focusing light.} The orientation of each mirror in the array is managed by applying a specific bit pattern to the DMD, with each mirror's angle determined by a corresponding bit: `1' causes reflection towards the projector, while `0' directs it away. The refresh rate denotes the frequency at which the entire mirror array's pattern can be updated, essentially defining the \textit{baud rate} of the signal transmitted.

As discussed in Section~\ref{sec:OOK} earlier, we group an $M \times M$ square array of mirrors into a single block for OOK control. Furthermore, to mitigate inter-channel light interference, we also maintain unused blocks adjacent to each active block, which further  the overall number of available channels in our VLC system.

The receiver employs a DVXplorer Lite event camera~\cite{xplorelite}, featuring a spatial resolution of 320 x 240 pixels. It boasts an event timestamp resolution of 1 $\mu$s, which theoretically ensures that it does not limit the refresh rates of the DMD in question. Nonetheless, as previously discussed and demonstrated in Section~\ref{subsec:eventLoss}, the event camera faces challenges in processing all events promptly at higher DMD refresh rates, leading to event loss. Consequently, the event camera effectively becomes a limiting factor for the DMD refresh rate to ensure reliable communication. This issue and its implications will be further analyzed in Section~\ref{s:result}.

\noindent \textbf{Experimental setup:}\footnote{Our experiments don't involve human subjects and have no
ethical concerns. We will release our hardware design, codes and dataset to
the public.} Considering the DLP4500's ability to project mirror reflections across a broad field of view, we designed our experiments such that the event camera is positioned to face a wall, capturing the multi-channel DMD data projections displayed on this surface (see Fig.~\ref{fig:realsetup}). This arrangement eliminates the need for additional modifications to either the DMD or the event camera, preserving the integrity of our experiments aimed at evaluating the data rate and BER performance of our proposed methodologies. In fact, this setup has numerous practical applications, including the distribution of data to multiple receivers within a room. In this scenario, the receivers are oriented towards a wall or ceiling upon which a central DMD projects the data. Potential applications include \yx{transmitting visual information and VLC data simultaneously in tourist venues, restaurants, and historical buildings, as Fig.\ref{fig:application} shows.}

In this experimental configuration, the DMD is stationed at a fixed distance of 1m from the wall. Conversely, the event camera is positioned at varying distances and angles relative to the wall to investigate the impact of distance and orientation on VLC communication efficacy. Additionally, this setup facilitates adjustments in room lighting and the projection surface material on the wall, accommodating a range of materials from standard wall finishes to fabric screens.

For individual channel data transmission, we employ a straightforward format detailed below. Each packet begins with an STX (Start of Text, \emph{0b01010101}) marker (a.k.a preamble), aiding in packet delineation and synchronization. Subsequently, the ASCII-encoded payload follows. Our default transmitted payload is the word `good', represented in as \emph{0b01100111 01101111 01101111 01100100}. The packet ends with an ETX (End of Text, \emph{0b00001111}) marker. BER calculations are performed on the payloads of packets with accurately received STX and ETX markers.

\subsection{Results}
\label{s:result}

\subsubsection{Maximum Data Rate.}
Recall that the data rate of our proposed multi-channel VLC system can be linearly scaled with the number of concurrent channels, limited by both the signal-to-noise ratio (SNR) and inter-channel interference. For instance, a too-small channel block size might result in an SNR too low for the event camera to discern light changes. Similarly, overly close placement of channels within the DMD mirror array could lead to significant lighting interference, preventing the event camera from generating reliable independent event streams for each channel. Through extensive evaluations, we determined a minimum channel size of $8 \times 8$ ($M=8$; 64 mirrors in total per channel) is necessary for adequate SNR, and spacing of one channel is required between two active channels to minimize inter-channel interference. Consequently, we can support a total of $N = 1,995$ channels, organized in a $57 \times 35$ array across the DMD micro-mirror array. During the calibration phase of the experiment, we align the DMD channels with the event camera pixels by transmitting the channel number in the payload at a low refresh rate of 588Hz.

The maximum data rate achievable with this configuration is essentially 1,995 times the usable DMD refresh rate. However, as illustrated in Fig.~\ref{fig:low-vs-high}, a higher refresh rate leads to increased event loss rates in the event camera, potentially resulting in a higher BER. Consequently, the usable refresh and hence the maximum achievable data rate, would be constrained by the acceptable BER threshold.

\begin{figure}[htp!]
    
    \begin{subfigure}[b]{0.49\columnwidth}
    \centering
    \includegraphics[width=\textwidth]{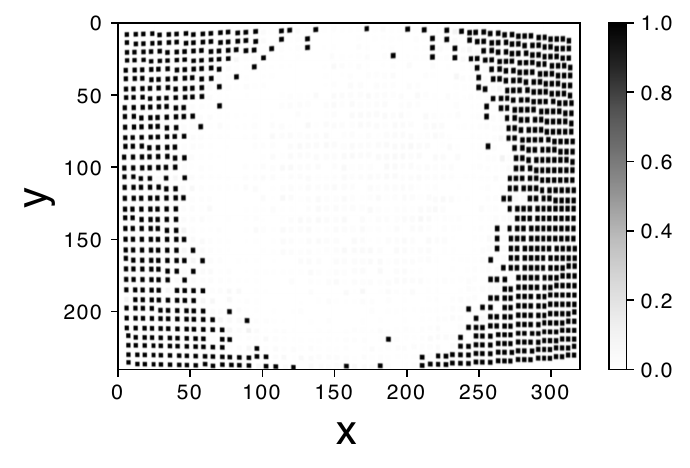}
    \caption{Single refresh rate. }
    \end{subfigure}
    \hfill
    \begin{subfigure}[b]{0.49\columnwidth}
    \centering
    \includegraphics[width=\textwidth]{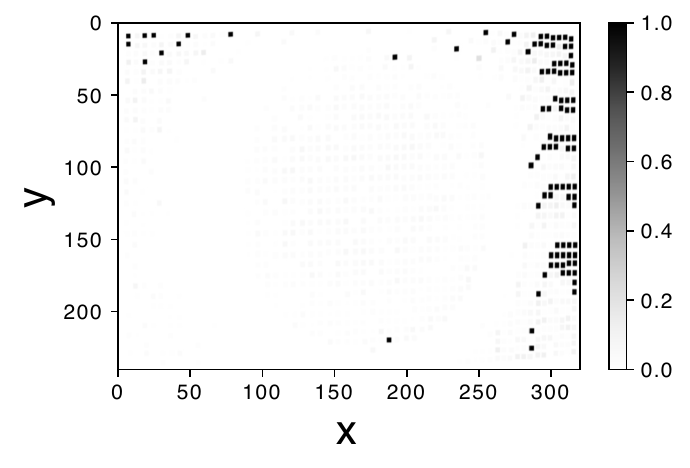}
    \caption{Dual refresh rate. }
    \end{subfigure}
    \caption{Spatial distributions of BERs. Single refresh rate: $f_c = f_p = 2,000 Hz$ and dual refresh rate: $f_c=2,000Hz, f_p=1,000Hz$.}
    \label{fig:ber_map}
\end{figure}

As explained in Sec.~\ref{sec:dual}, we can improve the overall BER by selecting lower refresh rates for the peripheral channels compared to the central channels. This motivation is demonstrated in Fig.~\ref{fig:ber_map}, which shows that the peripheral channels suffer from very high BER when a DMD refresh rate of 2,000 Hz is used for all channels, but they are brought under control using the proposed dual refresh rate policy with lower refresh rates used for the peripheral channels.

To determine the maximum achievable data rate, we assess both single and dual refresh rate policies. Given that \textit{relative time decoding} outperforms \textit{absolute time decoding}—a detail further elaborated in Section~\ref{subsubsec:tatVsErt} below —we utilize this method to achieve the highest data rate.  For the single refresh rate policy, the upper bound of data rate is 1,995 times the refresh rate. In contrast, the dual refresh rate policy yields a data rate of $1,995 \times (C \times f_c + (1-C) \times f_p)$, where 
C represents the proportion of total channels designated as central; $f_c$ and $f_p$ are the refresh rates for central and peripheral channels, respectively. The value of $C$ is determined by the radius $d$ of circle discussed in Section~\ref{sec:dual} earlier, which is empirically chosen as $d = 15$. This
results in 709 central and 1,286 peripheral channels.

\begin{figure}[htp!]
    \centering
    \includegraphics[width=0.8\columnwidth]{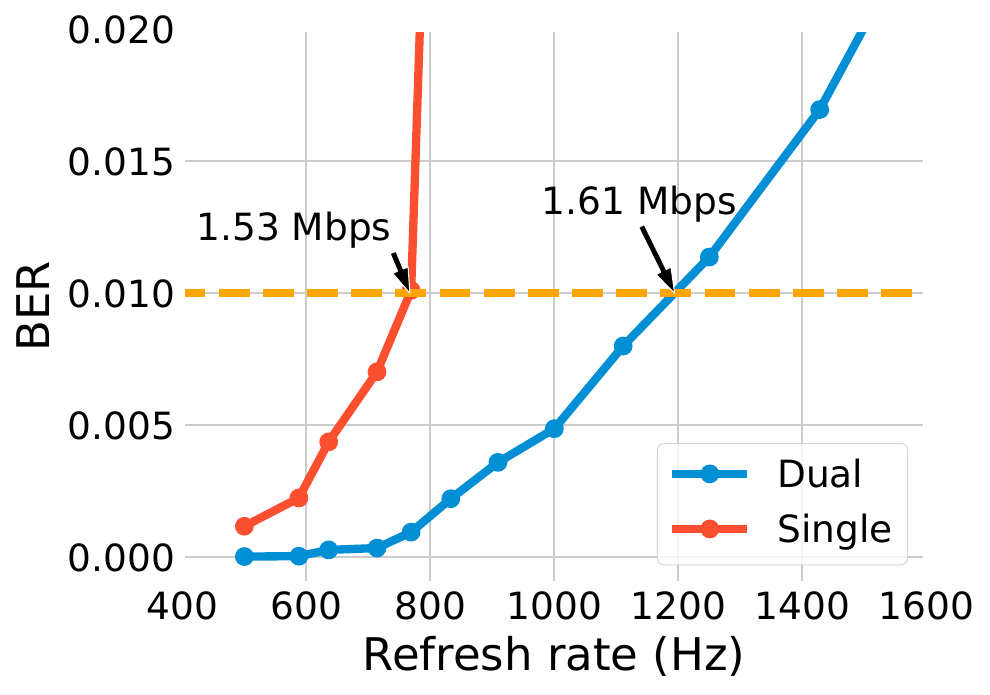}
    \caption{BER as a function of DMD refresh rate. The x-axis shows the refresh rate for central channels ($f_c$). For single rate, $f_p=f_c$. For dual rate, $f_p=\frac{f_c}{2}$}
    \label{fig:refreshrate}
\end{figure}

BER is calculated over payloads from 30 packets whose STX and ETX were received correctly. Thus, BER for each channel is calculated over $30 \times 32 = 960$ bits. Aiming for an average BER (over all channels) of 1\%, Fig.~\ref{fig:refreshrate} demonstrates that our proposed passive VLC system can achieve a data rate of 1.53 Mbps under a single refresh rate policy, surpassing the 1 Mbps threshold for passive VLC. Implementing dual refresh rates further enhances performance, reaching 1.61 Mbps and validating the dual refresh rate policy's efficacy. This achievement is noteworthy as it establishes a new benchmark for passive VLC, significantly exceeding the prior record of 100 kbps~\cite{xu2022exploiting}.

\subsubsection{Absolute vs. Relative Time Decoding}
\label{subsubsec:tatVsErt}

\begin{figure}[htp!]
    \centering
    \includegraphics[width=0.8\columnwidth]{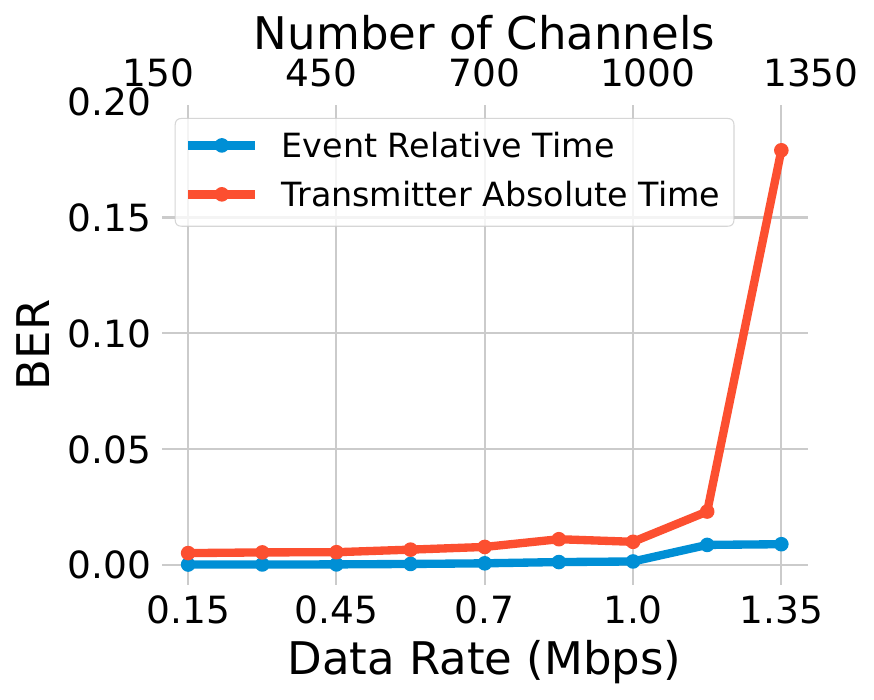}
    \caption{Comparison between Transmitter Absolute Time and Event Relative Time-based decoding.}
    \label{fig:absolute-relative}
\end{figure}

While keeping the refresh rate fixed to 1,000 Hz for all channels, we vary the number of channels to observe and compare the BER for both absolute and relative time decoding. The axis of Fig.~\ref{fig:absolute-relative} shows the number of channels and their corresponding data rates, while the y-axis reports the BER obtained for both coding schemes. The experiments demonstrate that the proposed event relative time-based approach outperforms conventional transmitter absolute time-based method, particularly when a larger number of channels are involved. For instance, with a data rate of 1.35 Mbps, the event relative time-based approach can achieve a BER of 0.88\% whereas the BER of the transmitter absolute time-based approach is 17.9\%. This superiority can be attributed to the robustness of the event relative time method in handling asynchronous time stamps and potential delays associated with a higher event count. In contrast, the transmitter absolute time-based method utilizes precise time stamps to ascertain the decoded bits. However, this approach is notably susceptible to inaccuracies in time stamps, which can arise due to the process queue within the MCU.

\begin{figure*}
\begin{subfigure}[b]{0.24\textwidth}
    \centering
    \includegraphics[width=\textwidth]{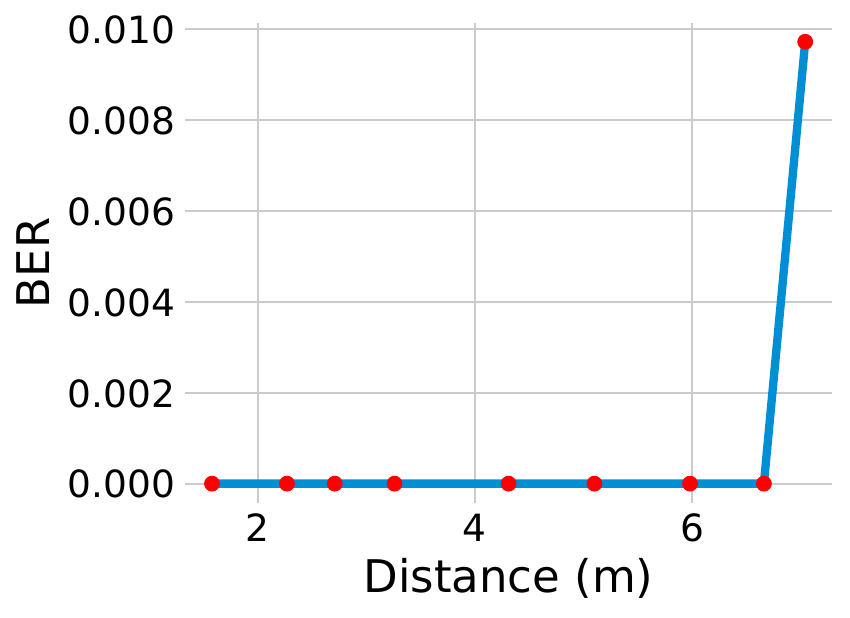}
    \caption{BER vs Distance.}
\end{subfigure}
\hfill
\begin{subfigure}[b]{0.24\textwidth}
    \centering
    \includegraphics[width=\textwidth]{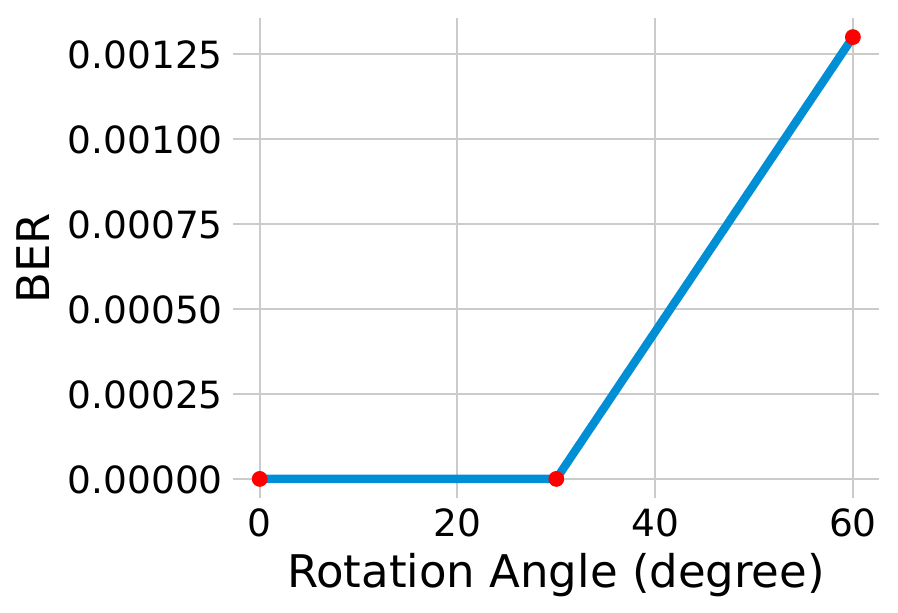}
    \caption{BER vs Rotation Angle.}
\end{subfigure}
\hfill
\begin{subfigure}[b]{0.24\textwidth}
    \centering
    \includegraphics[width=\textwidth]{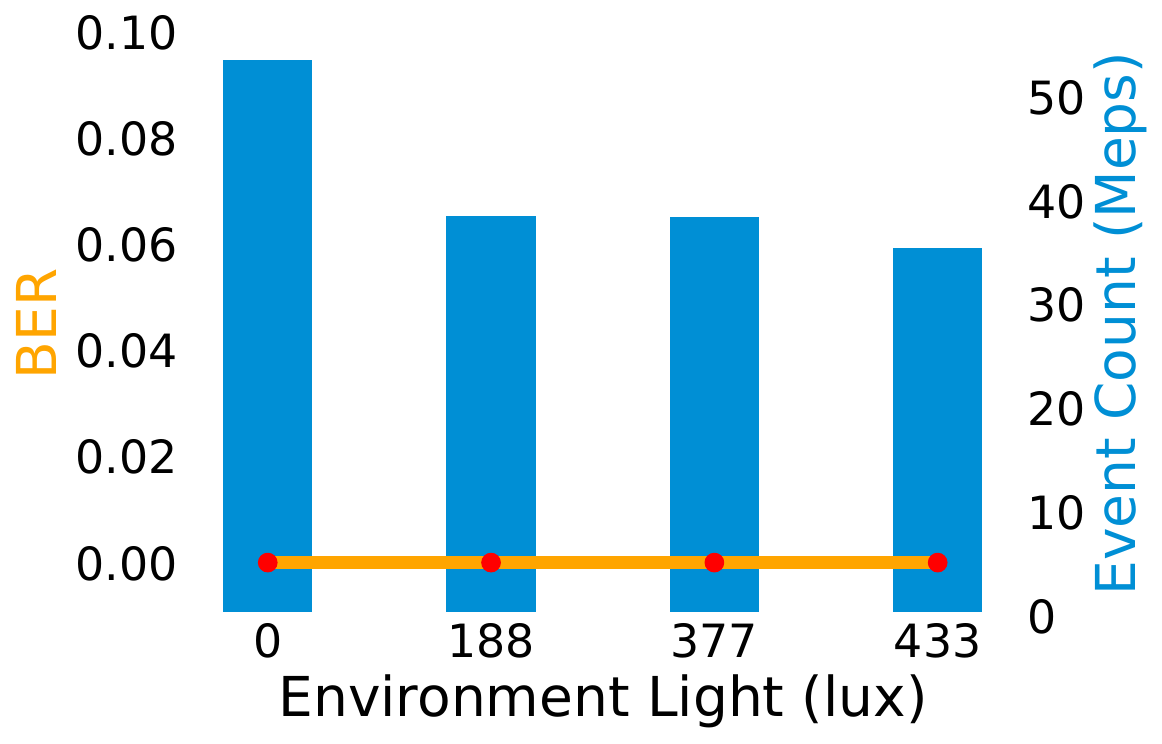}
    \caption{BER vs Environment Light.}
\end{subfigure}
\hfill
\begin{subfigure}[b]{0.24\textwidth}
    \centering
    \includegraphics[width=\textwidth]{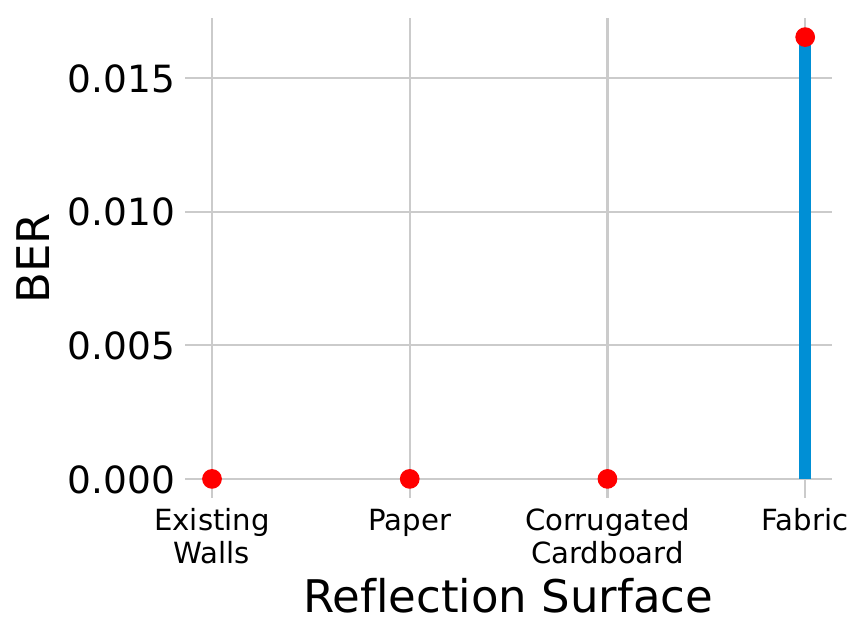}
    \caption{BER vs Reflection Surface.}
\end{subfigure}
\caption{The impact of distance, rotation angle, environment light and reflection surface.}
\label{fig:result-distance}
\end{figure*}

\subsubsection{Impact of Distance.}
To study the effect of distance,
the DMD is pointing to the wall while the distance is 1 meter initially. Then, we change the distance between the wall and the receiver. At each distance, we collect the events data for 5 seconds. 
Fig.~\ref{fig:result-distance}(a) shows that the BER increase to around $10^{-2}$ after six meters. As expected, when the distance increases, the SNR will decrease, which causes the errors in both channel mapping and event decoding. 
Our additional experiment validated, when a single channel is utilized (i.e., all the mirrors in the DMD share the same state), \cname functions similarly to the Photo-Link, with the transmission range extending to approximately 20 meters resulted from the increasing of SNR (by trading-off the data rates since we have one channel, i.e., $N = 1$, only here).

\subsubsection{Impact of Orientation.}
\begin{figure}[htp!]
    \centering
    \includegraphics[width=\columnwidth]{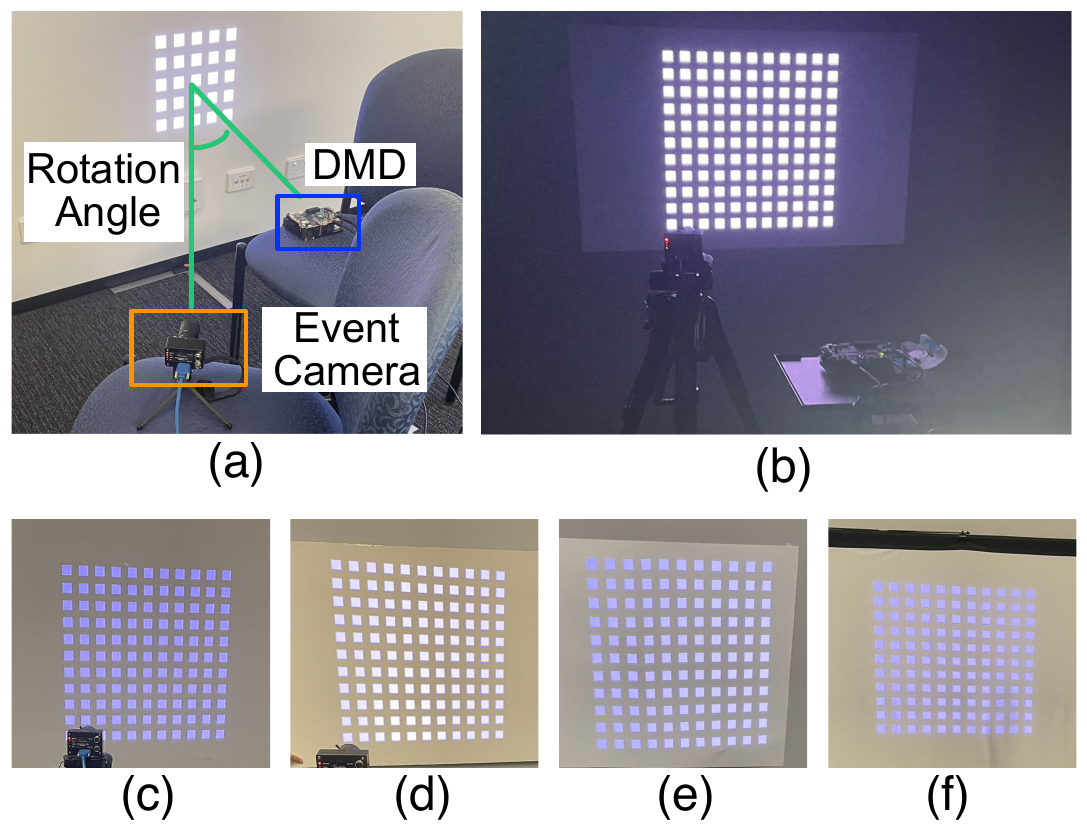}
    \caption{(a) Rotation angle (b) Environment light (c)-(f) Different reflective surfaces: existing wall (c), white paper (d), corrugated cardboard (e), and white fabric (f)}
    \label{fig:rotation}
\end{figure}

In this experiment, we set the DMD to perpendicular to the wall initially before rotating the receiver to have different orientation angles, i.e., 0, 30, 60 degrees to the perpendicular direction, as shown in Fig.~\ref{fig:rotation}(a). The results are shown in Fig.~\ref{fig:result-distance}(b). \yx{When the receiver is at a 60-degree angle, the perspective effect from the lens causes more channels to overlap with each other, resulting in inter-channel interference and increased BER. It is important to note that all lens-based camera systems are susceptible to similar issues. However, increasing the spacing between channels can help reduce overlapping effects and expand the range of workable angles.}

\subsubsection{Impact of Environment Light.}
To study how environment light affects the performance of \cname, we control the environment light's intensity by turning on/off the lights in the experiment room. 
Fig.~\ref{fig:rotation}(b) presents the scenario when the environment light intensity is 0 lux. Fig.~\ref{fig:result-distance}(c) shows the performance of \cname under four different levels of environment light intensity. The figure shows that the BERs are zeros for all four scenarios; however, the numbers of generated events are different due to the duplicate events discussed in Section~\ref{subsubsec:eventCameraPrinciples} earlier.

For example, when the room is dark (0 lux), the change in light intensity ($\Delta$) caused by controlling the states of the DMD is larger than when the environment light intensity is 433 lux. Therefore, there are more events produced, exceeding 50 million events per second (Meps), when the environment light intensity is 0 lux compared to when it is 433 lux, which results in less than 40 Meps. This justifies the necessity of duplicate event removal (see Section \ref{subsubsec:eventCameraPrinciples}) in our design. \yx{However, when the environmental light intensity is excessively high, such as direct sunlight exceeding 10,000 lux, the relative light intensity change caused by toggling mirrors is too weak, resulting in a low SNR. Consequently, the system will be affected severely since the event camera will be unable to emit any valid events.}

\subsubsection{Impact of the Material of Reflection Surface.}
We evaluate how the reflection surface will affect the performance of \cname. Here, 
we covered the existing wall with different materials, which include printing paper, corrugated cardboard that is used for packaging, and white fabric, as depicted in Fig.~\ref{fig:rotation}(d-f). The results are shown in the Fig.~\ref{fig:result-distance}(d). Apart from the white fabric, the BERs for the other three are all zeros. We observe there are a lot of small folds on the top of the white fabric, which create uneven local reflections and inter-channel interference, whereas the other three items have relatively flat surfaces. \yx{Choosing a relatively flat plane can provide better performance, and vice versa.}

\section{related work}
\label{sec:related}

Compared to active VLC, which modulates the light source directly~\cite{cui2021radioinlight,carver2021amphilight,ye2023vlc,cui2020sniffing,li2015real,ye2021spiderweb,carver2022air}, data rates in passive VLC are significantly lower. To achieve passive VLC, various smart surfaces have been explored for modulating light propagation features between the light source and the receiver, including switchable glass for blocking, LCD shutters for twisting, and DMD for reflection. Switchable glass-based passive VLC, alternating between opaque and transparent states, supports up to \textbf{33.33 bps}~\cite{hu2020nlc}. LCD shutters can twist linearly polarized light with varying voltages, achieving around \textbf{1 kbps} as reported by Ghiasi et al.~\cite{ghiasi2021principled}. While LC shutter stacking and backscattering \cite{ghiasi2023spectralux, xu2022low, wu2020turboboosting} can slightly increase LCD shutter-based VLC data rates, the inherent transition times of switchable glasses and LCD shutters pose a substantial bottleneck. PhotoLink\cite{xu2022exploiting} overcomed this bottleneck using a DMD as a super fast programmable reflector to reach \textbf{100 kbps} in passive VLC.
To our knowledge, \cname is the first implementation of multi-channel passive VLC communication employing a DMD as the transmitter and an event camera as the receiver, breaking the \textbf{1 Mbps barrier}. Unlike high-speed cameras~\cite{iwase2014improving}, which inefficiently capture every pixel in each frame regardless of light signal changes, leading to excess data processing, storage needs, and higher power consumption, event cameras offer a more efficient solution for passive VLC, particularly in applications with low-cost Internet of Things sensors.

The event camera has been explored across numerous domains, including Simultaneous Localization And Mapping (SLAM)~\cite{gallego2017event}, gesture recognition~\cite{amir2017low,wang2019ev}, depth estimation~\cite{zhu2019unsupervised,muglikar2021event}, object tracking~\cite{chen2020end}, and image enhancement~\cite{sun2023event,rebecq2019high}. Its application in VLC scenarios is a recent development. Shen et al.~\cite{shen2018vehicular} pioneered its use as a receiver in vehicular VLC, showcasing its effectiveness in mobile environments with a \textbf{16 kbps} data rate. Wang et al.~\cite{wang2022smart} expanded on this by employing event cameras in a one-pixel camera setup, achieving \textbf{4 kbps} indoors at 0.3m and \textbf{500 bps} outdoors at 100m. Another novel approach is the spike camera, introduced by Huang et al.~\cite{huang20221000}, which uses synchronization different to event cameras. Xu et al.\cite{xu2023visible} implemented this technology in a six-channel light communication system with LCD shutters, reaching a \textbf{4.8 kbps} data rate.

\textcolor{revised}{Screen-to-camera (S2C) communication is one special case of modulating light sources. Unlike blinking light bulbs, screens can establish more fine-grained light connections. Cameras are mostly used to capture information from multiple pixels, with the rolling shutter effect providing a higher sampling rate~\cite{danakis2012using}. To maintain the original display functions of screens and avoid disruption, extensive research has been conducted on how to conceal communication while preserving normal display performance.
InFrame~\cite{wang2014inframe} proposed the use of adjacent frames to construct complementary frames. Due to the human eye's low-pass filter effect, the complementary pixel values cannot be detected by the naked eye but can be captured by cameras. Additionally, in HiLight~\cite{li2015real}, the information is encoded in the transparency values of the pixels, reducing the flickering effect caused by complementary frames. Similarly, InFrame++~\cite{wang2015inframe++} designed smaller and continuous complementary cells to achieve similar goals. In the case of ChromaCode~\cite{zhang2018chromacode}, the bit stream is encoded through color changing. Specifically, ChromaCode selects color transitions that are suitable for the human eye, enabling a flicker-free effect. Apart from using screens exclusively, AIRCODE\cite{qian2021aircode} separates the control and data frames into audio and video channels. The works mentioned above share a common feature: they rely on human-designed algorithms to extract embedded information. In contrast, DeepLight~\cite{tran2021deeplight} leverages the advantages of neural networks to extract data bits from images without prior knowledge, such as image size and orientation, making the entire system more practical. Unlike the aforementioned S2C communications, \cname focuses on the VLC communication with the DMD flashing at high speed, which has the potential to incorporate display functions in future works. }

\section{discussions and limitations}
\label{sec:limitation}

\paragraph{Moving Scenarios.}

\yx{\cname has the potential to operate effectively not only in static settings but also in dynamic scenarios such as vehicle-to-vehicle communication. The core principle of \cname, which involves using different mirrors on the DMD for data transmission and an event-based vision camera for decoding, would remain consistent even in moving situations. For example, the light channels could be distributed across the covers of car lights. However, the primary challenge in such dynamic scenarios would be the constantly changing channel positions, as both the transmitter and receiver could be in motion. This would require advanced tracking and alignment techniques to maintain a stable, low-latency communication link and can be a future work.}

\textit{Practicality, \textcolor{revised}{Cost and Power Consumption}.}
\yx{
While the applications of \cname align with conventional VLC, the DMD chip's microscopic mirrors (typically 7-10 $\mu$m in size) necessitate an intermediate surface to expand light channels. Direct reception from the DMD would result in channel mixing, so visible light channels can be instead projected onto surfaces like light covers, walls, floors, and ceilings.
Cost of \cname can be reduced significantly using lower-resolution DMD chips. For example, a low-resolution DMD chip can cost as little as \$40~\cite{xu2022low}. \textcolor{revised}{
As for the power consumption, the event camera consumes approximately 5 to 14 mW~\cite{gallego2020event}. For the transmitter, the power consumption is around 442mW, but it can be reduced to a minimal 45mW by using smaller number of mirrors~\cite{xu2022exploiting}. The LED light source conjugated in the DMD evaluation module used in the experiments consumes around 15 W~\cite{dmd}.
}
}

\textit{Alternatives of the Light Source.}
This system employs an LED light source with a DMD evaluation module, but it's adaptable to various light types, including infrared, sunlight, and laser. Each alternative offers unique benefits: infrared enables invisible communication, making it a better option for communication that would not interfere with the visual environment. And using the laser light could enhance the system's workable range, allowing for communication over longer distances. These options could broaden the applications and enhance the performance of DMD-based passive VLC systems in diverse settings.

\textcolor{revised}{
\textit{More than OOK Modulations}. In this paper, we implemented On-Off Keying (OOK) modulation on the transmitter side. However, several other modulation schemes can also be applied. For instance, Manchester encoding uses differential changes between ON and OFF states to represent different symbols and exhibits less flickering compared to OOK, especially when transmitting long sequences of repeated bits. In addition, Frequency Shift Keying (FSK) is more robust to dynamic ambient light interference, to which OOK is vulnerable. Lastly, OOK's data rate can be considered the upper bound for understanding the system's throughput capability.}

\begin{figure}[htp!]
    \centering
    \includegraphics[width=0.9\linewidth]{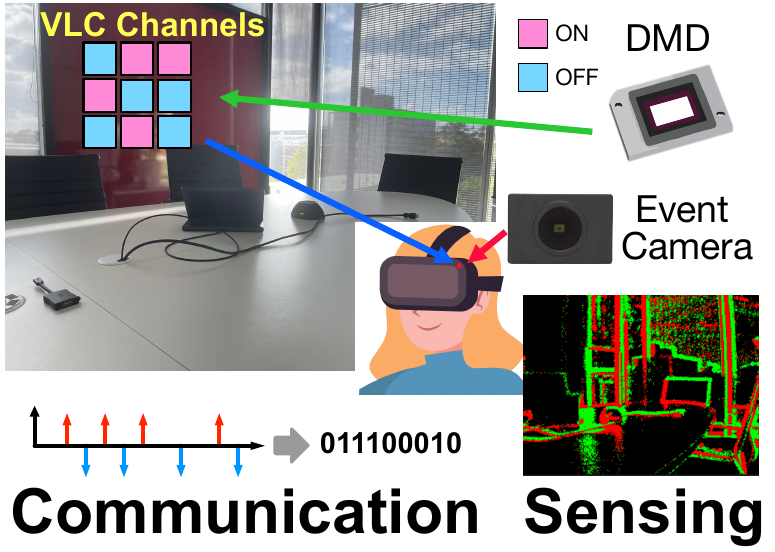}
    \caption{In AR/VR applications, event cameras integrated into head trackers can simultaneously function as VLC receivers for \cname-powered communications and sensors for head pose estimation~\cite{gallego2017event}.\protect\footnotemark
    }
    \label{fig:motivation}
\end{figure}

\textcolor{revised}{
\textit{The Potential Applications.} In the proposed system, light channels are redirected to another surface, such as ceiling, walls or ground. A multitude of receivers from different locations can face the same surface to receive the VLC data simultaneously. Thus, with planned geometry, the system can overcome conventional line-of-sight limitations of VLC, enabling a wider and more flexible communication area. \cname is also suitable for secure mass data dissemination  in secret enclosed spaces, such as at military locations where the troops gather to receive highly confidential updates. Since light cannot penetrate walls, it provides an inherent layer of security against external eavesdropping. Furthermore, event cameras used in vision-based sensing applications, can also function as receivers for enabling communication~\cite{nishar2024joint}. For instance, in AR/VR applications~\cite{gallego2017event}, event cameras embedded in headsets for estimating the 6-DOF head movements could simultaneously capture the VLC channels projected on surrounding surfaces, as illustrated in Fig.~\ref{fig:motivation}. This dual functionality paves the way for potential integration of visible light sensing and communication capabilities in metaverse applications.
}
\footnotetext{This figure has been designed using assets from Freepik.com}

\section{conclusion}
\label{sec:conclusion}

This paper introduces \cname, a novel VLC system that enables multi-channel communications. We propose an innovative event-based decoding algorithm specifically designed for the unique characteristics of event generation. Our multi-channel approach achieves a maximum data rate of 1.6 Mbps, representing a 16-fold improvement over PhotoLink~\cite{xu2022exploiting}. While \cname remains constrained by the receiver's physical limitations, it demonstrates potential \textcolor{revised}{ directions towards} significantly higher data rates in passive VLC systems.

\begin{acks}
We are grateful for the insightful comments and keen observations from the anonymous shepherd and reviewers, which have significantly improved this paper. This work was partly supported by the Australian Research Council Discovery Project DP210100904, UNSW Scientia PhD Scholarship Scheme and CSIRO Data61 PhD Scholarship Program.
\end{acks}

\balance
\bibliographystyle{ACM-Reference-Format}
\bibliography{reference}

\end{document}